# Lead-free piezoelectric perovskites for arterial-pulse e-skin: from configurational complexity to equivariant machine-learning potentials

*A Review and Perspective*

**Sanaa Ismail[1,2], Hassan M. E. Azzazy[1]**

[1] Department of Chemistry, School of Sciences and Engineering, The American University in Cairo, AUC Avenue, P.O. Box 74, New Cairo 11835, Egypt

[2] Department of Materials Science and Engineering, The Pennsylvania State University, University Park, PA 16802, USA

## Abstract

Continuous, non-invasive monitoring of the arterial pulse is a clinical priority for cardiovascular disease, the leading cause of global mortality. Flexible piezoelectric electronic skins can transduce the 1–10 kPa pressure wave into a self-powered voltage, but the best-performing piezoceramics are lead-based, and their toxicity is incompatible with skin contact and with tightening RoHS/REACH regulation. Among lead-free alternatives, the $BaTiO_3$-based solid solution BZT–BCT reaches $d_{33} \approx 620$ pC/N near its tricritical morphotropic phase boundary, rivalling soft PZT while remaining biocompatible. Exploiting this in a wearable confronts a sensitivity–flexibility paradox and three computational walls: the combinatorial explosion of atomic configurations in a disordered solid solution, the band-gap error of affordable density-functional approximations, which corrupts leakage and insulation estimates, and the 0 K nature of standard calculations against a 310 K operating temperature. We review lead-free piezoelectrics, morphotropic-boundary physics and fabricated flexible devices, then argue that equivariant machine-learning interatomic potentials — coupled to a tiered functional hierarchy and finite-temperature lattice dynamics — can survey the full configurational ensemble at body temperature and close the gap to a clinically viable lead-free pulse sensor.

# 1. Introduction & The Clinical Mandate

## 1.1 The Critical Need for Flexible Arterial Pulse Monitoring

Cardiovascular diseases (CVDs) remain the foremost cause of death worldwide, accounting for an estimated 19.8 million deaths in 2022—about 32 % of all global mortality [1]. A defining feature of CVD management is that its most actionable diagnostic information—arterial blood pressure (BP), heart-rate variability, arterial stiffness, and the morphology of the pressure pulse wave—is intrinsically dynamic, varying from beat to beat and over longer physiological cycles [2], [3]. That variability is not measurement noise: short- and long-term BP variability carries prognostic information beyond mean BP [3], and aortic stiffness quantified as pulse wave velocity independently predicts cardiovascular events and all-cause mortality [4]. Yet the clinical standard of care, the intermittent oscillometric cuff, samples this continuum only sparsely; such isolated readings neither reconstruct the out-of-office BP profile on which current diagnosis and management are based [2] nor track the beat-to-beat dynamics that cuffless wearable devices are now being developed to resolve [5]. This mismatch between the dynamic nature of the signal and the static nature of its measurement is the clinical mandate motivating a shift from episodic readings toward continuous, non-invasive, wearable monitoring at the skin surface [6].

The radial and carotid pressure pulses that reach the epidermis are mechanically subtle: the transcutaneous pressure swing of a single heartbeat lies in the range of approximately 1–10 kPa [7]. Faithful transduction of this low-pressure window demands a sensor that is simultaneously (i) highly sensitive at low mechanical load, (ii) fast enough to resolve intra-beat features such as the dicrotic notch, which occupy a few tens of milliseconds within the 600–1000 ms cardiac cycle of a resting adult [8], and (iii) mechanically compliant enough to conform intimately to the curved, deformable surface of skin. Rigid metallic and silicon transducers fail the third criterion: their high flexural modulus imposes a mechanical-impedance mismatch with soft tissue, whose in vivo indentation modulus is of order 10–100 kPa, although reported values depend strongly on skin layer, hydration and measurement method [9], producing motion artefacts, unstable contact, and signal degradation during natural movement [6]. Optical wearables based on photoplethysmography (PPG) sidestep this mechanical mismatch but infer the pulse only indirectly from blood-volume changes and remain susceptible to motion and perfusion artefacts [8], so a conformal transducer that measures pulse pressure directly is preferred for high-fidelity waveform capture.

These constraints have converged on the paradigm of the flexible electronic skin (e-skin)—thin, conformal sensor arrays that mechanically emulate the epidermis while transducing physical stimuli into electrical signals [6], [10]. Among the available transduction modes (piezoresistive, capacitive, triboelectric, and piezoelectric), the piezoelectric mode is uniquely attractive for cardiovascular monitoring because it is intrinsically self-powered: the mechanical energy of the cardiac cycle is converted directly into a measurable open-circuit voltage ($V_{oc}$) without an external bias supply, removing the bulky energy-storage components that burden conventional wearables [7], [10]. The performance ceiling of such a device is therefore set, first and foremost, by the piezoelectric figure of merit of its active material—chiefly the longitudinal piezoelectric charge coefficient $d_{33}$ and the voltage coefficient $g_{33}$—which dictates how much electrical signal is generated per unit of arterial pressure; these figures of merit are defined and compared in §2.1.

## 1.2 The Regulatory & Biosafety Push for Lead-Free Piezoelectrics

For more than half a century the piezoelectric market has been dominated by lead zirconate titanate, $Pb(Zr,Ti)O_3$ (PZT) [11], whose morphotropic-phase-boundary compositions deliver $d_{33}$ values of 500–600 pC/N and remain the benchmark for high-end actuators and sensors [12]. This commercial supremacy is, however, fundamentally incompatible with the trajectory of both environmental regulation and wearable biomedicine, because PZT contains more than 60 wt% lead oxide [12]—and lead is a cumulative toxicant that is stored in bone and teeth and for which no level of exposure is known to be without harmful effects [13].

On the regulatory front, lead has been progressively restricted under the European Union Restriction of Hazardous Substances (RoHS) and Registration, Evaluation, Authorisation and Restriction of Chemicals (REACH) frameworks, together with the Waste Electrical and Electronic Equipment (WEEE) directive [14]. Lead in PZT piezoelectric ceramics has so far retained a time-limited exemption (RoHS Annex III—historically entry 7(c)-I, refined in 2025 into entry 7(c)-VI and currently set to expire on 31 December 2027), granted only because no adequate substitute yet exists; such exemptions are subject to periodic review and sunset, creating an explicit regulatory mandate to develop lead-free materials that can match PZT before they lapse [14], [15].

For skin-mounted and potentially implantable devices the imperative is sharper still, and here the biosafety argument supersedes the environmental one. A transducer in chronic, intimate contact with perspiring skin—or in direct contact with tissue—cannot, on precautionary grounds, risk leaching $Pb^{2+}$ ions, and neither device failure nor disposal should introduce lead into the body or the waste stream [13]. Lead-free piezoelectrics are therefore not merely a sustainable preference but a biocompatibility prerequisite for this application class [15], [16]. Among lead-free perovskite oxides, barium titanate ($BaTiO_3$, BTO)-based solid solutions are especially appealing because BTO shows good compatibility with normal cells such as fibroblasts and neurons and has itself been explored in biomedical settings, albeit with dose- and cell-type-dependent effects [17]. The pseudo-binary system $(1-x)Ba(Zr_{0.2}Ti_{0.8})O_3–x(Ba_{0.7}Ca_{0.3})TiO_3$ (BZT–BCT) is the standout candidate: at its morphotropic phase boundary it attains $d_{33}$ ≈ 620 pC/N, reaching parity with soft PZT in a conventionally sintered ceramic and far exceeding pure BTO (≈150 pC/N) [10] and base alkali-niobate (K,Na)$NbO_3$ (KNN, ≈80–110 pC/N in base form; only heavily engineered KNN reaches ~400–700 pC/N, and then at reduced $T_c$ — see §2.2) [12], [18], [19]. BZT–BCT thus uniquely couples a clinically sufficient sensitivity with the biocompatibility demanded by skin-contact operation.

### 1.3 The Central Paradox: Giant Sensitivity vs. Mechanical Compliance in Ceramics

The promise of BZT–BCT exposes the central materials dilemma around which this review is organized. The very property responsible for its giant sensitivity—a soft, easily rotated polarization arising near a tricritical morphotropic phase boundary—is a bulk, crystalline-ceramic phenomenon, whereas the property demanded by the e-skin application—mechanical compliance—is antithetical to the dense, stiff, brittle nature of perovskite-oxide ceramics. Sintered BZT–BCT has an elastic modulus of order 100 GPa (Table 1), some six orders of magnitude stiffer than skin [9]; in monolithic form it can neither conform to tissue nor survive the flexion a wearable must tolerate.

The conventional routes to reconcile this conflict—dispersing the ceramic as nanoparticles, nanowires, or nanofibers within a flexible polymer matrix, or introducing engineered porosity to lower the effective stiffness—all succeed mechanically at a direct cost to the electrical signal. Diluting the active ceramic into a low-permittivity polymer, or replacing load-bearing ceramic with void, reduces both the effective $d_{33}$ and the poled volume fraction, so device-level sensitivity falls well below the intrinsic single-crystal limit [10], [20]. This is the sensitivity–flexibility paradox: every micro-structural strategy that buys compliance tends to spend the very piezoelectric response that justified choosing BZT–BCT. Resolving it is not a matter of a single fabrication trick but of understanding—and ultimately optimizing—the underlying structure–property relationships (phase competition, polarization anisotropy, and electromechanical coupling) at the atomic scale, which is precisely where first-principles computation becomes indispensable.

### 1.4 Scope, Thesis Statement & Article Structure

**Scope and positioning.** This article is a combined Review and Perspective. Sections 2–4 critically review established knowledge: the lead-free materials landscape, the physics of the morphotropic phase boundary, the micro-structural engineering used to confer flexibility, the state of the art in fabricated lead-free piezoelectric pulse sensors, and the limits of conventional first-principles methods. Sections 5–7 then advance a forward-looking perspective on how the resulting computational bottleneck can be overcome. We deliberately restrict the scope to lead-free, $BaTiO_3$-based piezoelectric perovskites operated as wearable mechanoreceptive (pulse/pressure) transducers; alkali-niobate and bismuth-based systems are treated comparatively rather than exhaustively, non-piezoelectric transduction modes are discussed only for context, and classical biosensors based on biochemical recognition lie outside our remit. Unlike prior reviews of perovskite sensors [10], [21]–[23], which primarily catalogue materials and devices, this article foregrounds the computational-discovery bottleneck and the equivariant-machine-learning route to overcome it.

***Terminology note.*** We adopt the term "piezoelectric e-skin pulse sensor" throughout for precision. Such devices fall under the broad wearable-biosensor umbrella, but—unlike a biosensor in the strict IUPAC sense, which couples a biological recognition element directly to a transducer [24]—they transduce a physical (mechanical) stimulus rather than a biochemical recognition event; this distinction is kept explicit to avoid the scope confusion common in the perovskite-sensor literature.

**Thesis statement.** Our central argument is that the decisive obstacle to the computational discovery of clinically viable lead-free e-skin materials is configurational complexity. A multi-component solid solution such as BZT–BCT presents an astronomically large number of inequivalent atomic arrangements on the perovskite A- and B-sites, which conventional 0 K density functional theory (DFT) can neither enumerate nor screen at tractable cost—and which it cannot, in any case, evaluate at the 310 K physiological temperature where the device must operate. We contend that equivariant graph-neural-network machine-learning interatomic potentials (MLIPs) are the enabling technology that closes this gap: by triaging the configurational space at near-DFT accuracy and granting access to finite-temperature thermodynamics, they transform the search from a biased sampling of a few hand-built structures into a systematic survey of the full configurational ensemble.

**Article structure.** Sections 2 and 3 build the materials case and then confront it with what has actually been fabricated, and Section 4 sets out the three walls on which conventional first-principles screening fails: the configurational explosion, the band-gap error that corrupts any leakage estimate, and the 0 K limitation. Section 5 introduces equivariant MLIPs and delimits precisely what they can and cannot predict, and Section 6 assembles them into a staged screening framework. Section 7 addresses the multiscale route from a computed coefficient to a device voltage and the open problem of the wet skin interface, before the conclusions of Section 8. Figure 1 sets out the whole argument in a single view, and the reader who wants the thesis before the detail may treat it as a map of what follows.

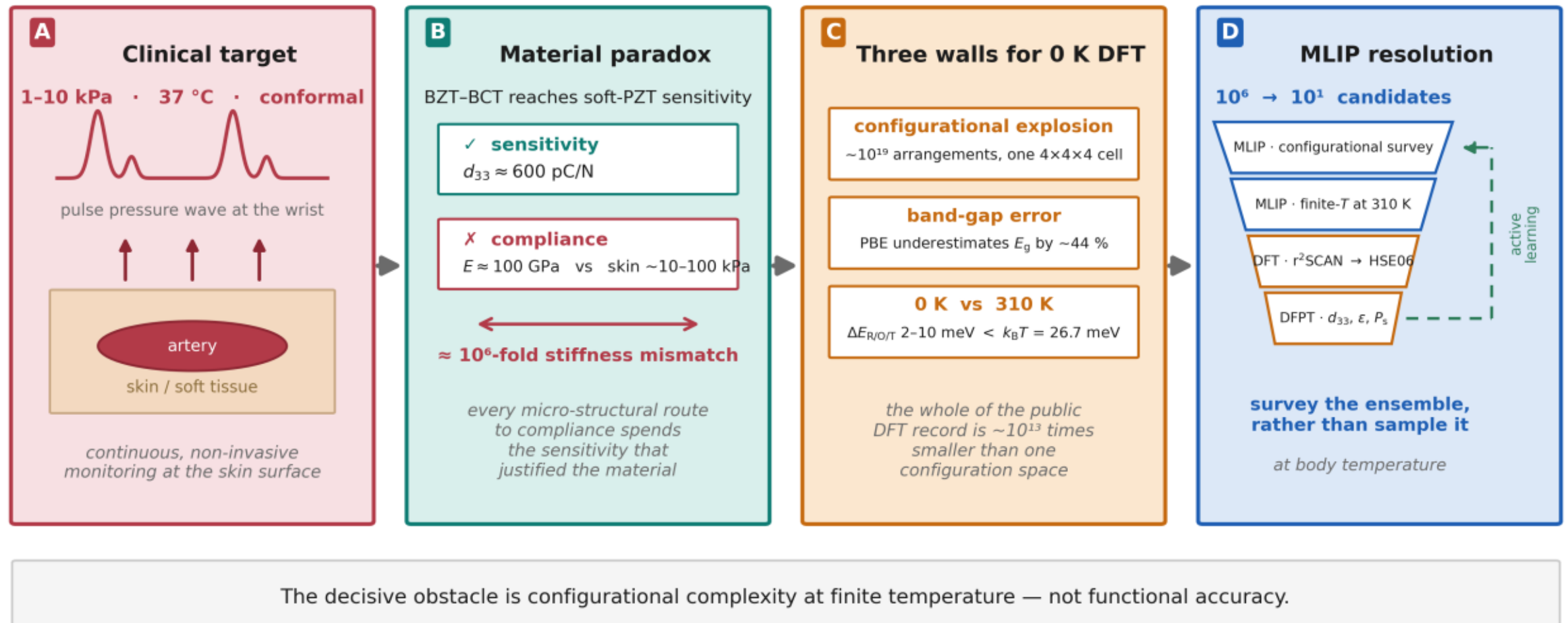

Figure 1. The argument of this Review and Perspective in one view. (a) The clinical target: continuous, non-invasive monitoring of the arterial pressure wave — a 1–10 kPa signal that must be transduced conformally, at body temperature, on deforming skin. (b) The material paradox: the lead-free solid solution BZT–BCT reaches soft-PZT sensitivity ($d_{33} \approx 600$ pC/N) but is some six orders of magnitude stiffer than skin, so every micro-structural route to compliance spends the very sensitivity that justified choosing it. (c) Three walls confront conventional 0 K density functional theory: the combinatorial explosion of dopant arrangements on the perovskite A- and B-sites, the systematic band-gap error of affordable functionals that corrupts any leakage estimate, and the near-degeneracy of the ferroelectric phases, whose separation is smaller than the thermal energy at 310 K. (d) The resolution advanced here: a fine-tuned equivariant machine-learning interatomic potential surveys the configurational ensemble and its finite-temperature behaviour, passing only a small set of survivors to density functional theory and density-functional perturbation theory, with an active-learning return path that returns new labels to the potential.

## 2. Material Landscape & The Sensitivity–Flexibility Paradox

### 2.1 Transduction Mechanisms & Figures of Merit

A wearable pulse sensor converts a mechanical pressure stimulus into a readable electrical signal through one of four dominant mechanisms (Figure 2): piezoresistive (stimulus changes resistance), capacitive (stimulus changes a gap or permittivity), triboelectric (contact electrification generates charge), and piezoelectric (stress generates bound surface charge) [21]–[23]. The first two are passive and require an external power supply and bias circuitry; the latter two are active and self-generating. For continuous, battery-free cardiovascular monitoring the piezoelectric mode is the most attractive single-material route, because the same crystal that senses the pulse also powers the readout, and—unlike the triboelectric mode—its output is governed by an intrinsic, tabulable material tensor rather than by interfacial contact conditions that are hard to predict from first principles [22], [23]. This is the mode on which the present review concentrates.

**Figures of merit.** Four figures of merit—the first three intrinsic to the material and defined here following the IEEE standard on piezoelectricity [25], the fourth a device-level quantity—determine how well a piezoelectric material serves as a low-pressure pulse sensor:

- $d_{33}$ — the longitudinal piezoelectric charge coefficient (pC/N): charge density generated per unit applied stress. It sets the short-circuit (charge) sensitivity and is the headline number for actuators and charge-mode sensors.
- $g_{33}$ — the piezoelectric voltage coefficient ($\times 10^{-3}$ V·m/N), related to $d_{33}$ by $g_{33} = d_{33} / (\varepsilon_0 \varepsilon_r)$. It sets the open-circuit voltage ($V_{oc}$) sensitivity and therefore the self-powered signal. Critically, because $g_{33}$ is inversely proportional to permittivity, a low-$d_{33}$ but low-permittivity polymer (PVDF, $g_{33} \approx 200$–$330 \times 10^{-3}$ V·m/N [26]) can rival a high-$d_{33}$ but high-permittivity ceramic (BZT–BCT, $g_{33} \approx 11$–$23 \times 10^{-3}$ V·m/N) in voltage output—a nuance that $d_{33}$ alone hides and that motivates ceramic–polymer composites (§2.4).
- $k_{33}$ and $k_p$ — the electromechanical coupling factor (dimensionless): the fraction of mechanical energy converted per cycle and the key efficiency metric for self-powered operation; $k_{33}$ (longitudinal) and $k_p$ (planar) are distinct modes.
- Sensitivity S — the device-level transfer slope (V·kPa$^{-1}$ or kPa$^{-1}$) over the physiological 1–10 kPa window, together with linearity, response time (target ≈ 20 ms [10]), and the workable pressure range.

A composite self-powered FOM, FOM = $d_{33} \cdot g_{33}$ (closely related to the coupling factor $k^2$), captures the energy-conversion capability that neither coefficient conveys alone, and is the metric energy-harvesting studies report [27]. Table 1 compiles these FOMs for the principal lead-free contenders alongside the PZT and relaxor benchmarks, and tabulates the product $d_{33}\cdot g_{33}$ explicitly, because it answers an objection the preceding paragraph invites: although PVDF has a far larger $g_{33}$, its very small $d_{33}$ caps the product, so on $d_{33}\cdot g_{33}$ BZT–BCT overlaps the soft-PZT benchmark and exceeds PVDF at the upper end. Three features stand out: (i) BZT–BCT attains soft-PZT-class $d_{33}$ in a conventionally sintered, randomly oriented bulk ceramic; (ii) the only lead-free entries that exceed it—textured $BaTiO_3$ and template-grain-grown KNN—do so through elaborate texturing routes that are difficult to reconcile with the compliant architectures of §2.4; and (iii) every high-$d_{33}$ ceramic is mechanically stiff (Young's modulus ~100 GPa), foreshadowing the paradox of §2.3–2.4. Reassuringly for the application, $BaTiO_3$-family sensors already demonstrate a workable range of 1–100 kPa with ~20 ms response—comfortably spanning the arterial-pulse window (Figure 3) [10].

Table 1. Comparison of piezoelectric figures of merit for lead-free candidates and lead-based benchmarks. Values are representative literature ranges; exact figures depend strongly on composition, poling, and processing. Entries marked (calc.) are computed from $g_{33} = d_{33}/(\varepsilon_0\varepsilon_r)$ with representative permittivities ($BaTiO_3$, $\varepsilon_r \approx 1400$–$1900$; base KNN, $\varepsilon_r \approx 400$–$500$; modified/textured KNN, $\varepsilon_r \approx 1300$–$2500$); the composite figure of merit $d_{33}\cdot g_{33}$ is evaluated from the tabulated ranges and quoted to two significant figures. Note that $d_{33}$ alone does not rank the candidates for a self-powered voltage sensor: PVDF, despite a $d_{33}$ roughly twenty times smaller than that of BZT–BCT, remains competitive on $d_{33}\cdot g_{33}$, while the two lead-free entries that exceed BZT–BCT in $d_{33}$ reach those values only through templated texturing. Polymer matrices (~1–4 GPa) sit ~2 orders of magnitude below the ceramics, and soft skin (~10–100 kPa) is six to seven orders below — the crux of the sensitivity–flexibility paradox.

| **Material (form)** | **$d_{33}$ (pC/N)** | **$g_{33}$ ($\times 10^{-3}$ V·m/N)** | **$d_{33}\cdot g_{33}$ ($\times 10^{-15}$ m²/N)** | **$k_{33}$ / $k_p$** | **$T_c$ (°C)** | **E (GPa)** | **Workable P (kPa) / response** | **Ref.** |
|---|---|---|---|---|---|---|---|---|
| **PZT-5H (soft, Pb)** | 500–620 | ≈20 | 10 000–12 400 | 0.65–0.75 | 195–230 | 60–70 | 1–65 / <30–65 ms | [10], [18] |
| **PMN-PT (single xtal, Pb)** | 1500–2500 | ≈30 | 45 000–75 000 | 0.90 | 130–170 | 15–25 | — (benchmark only) | [18] |
| **BZT–BCT 50BCT (bulk)** | ≈560–620 | ≈11–23 | 6 200–14 300 | 0.50–0.58 ($k_p$) | ≈93 | ≈100–130 | 1–100 / ≈20 ms | [18], [19], [28] |
| **$BaTiO_3$ (pure / textured)** | ≈145–190 / up to 788 | ≈9–15 (calc.) | 1 300–2 900 | 0.35–0.50 | ≈120 | 67–125 | 1–100 / ≈20 ms | [10], [20], [29] |
| **KNN ($K_{0.5}Na_{0.5}NbO_3$, base)** | ≈80–110 | ≈20–30 (calc.) | 1 600–3 300 | 0.30–0.45 | >400 | ≈100–120 | — / ~100 ms | [27], [30], [31] |
| **KNN (modified / textured)** | ≈344–700 | ≈15–30 (calc.) | 5 200–21 000 | ≈0.46 ($k_{31}$) | 200–510 (falls as $d_{33}$ rises) | ≈100–120 | harvester / — | [27], [31], [32] |
| **PVDF / P(VDF-TrFE)** | ≈20–33 | 200–330 | 4 000–10 900 | 0.12–0.20 | melt ~80 | 1–4 | 1–30 / 16–100 ms | [10], [26] |

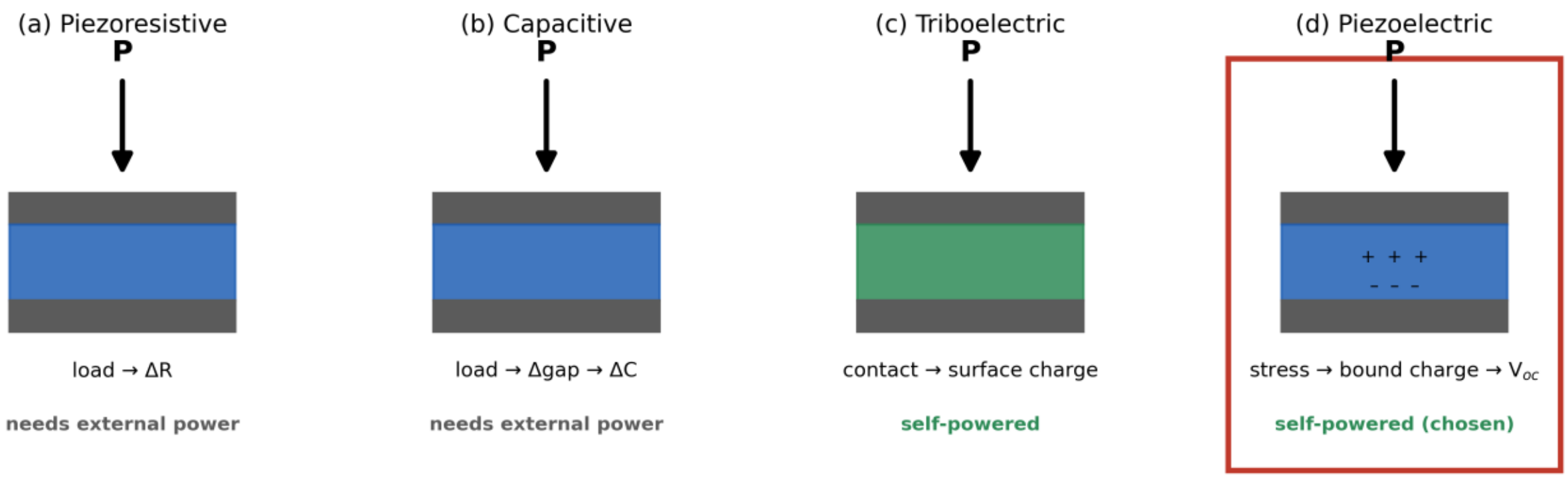

*Figure 2. The four pressure-transduction mechanisms. Only triboelectric (c) and piezoelectric (d) are self-generating; the piezoelectric mode (highlighted) is governed by an intrinsic, computable material tensor and is adopted here.*

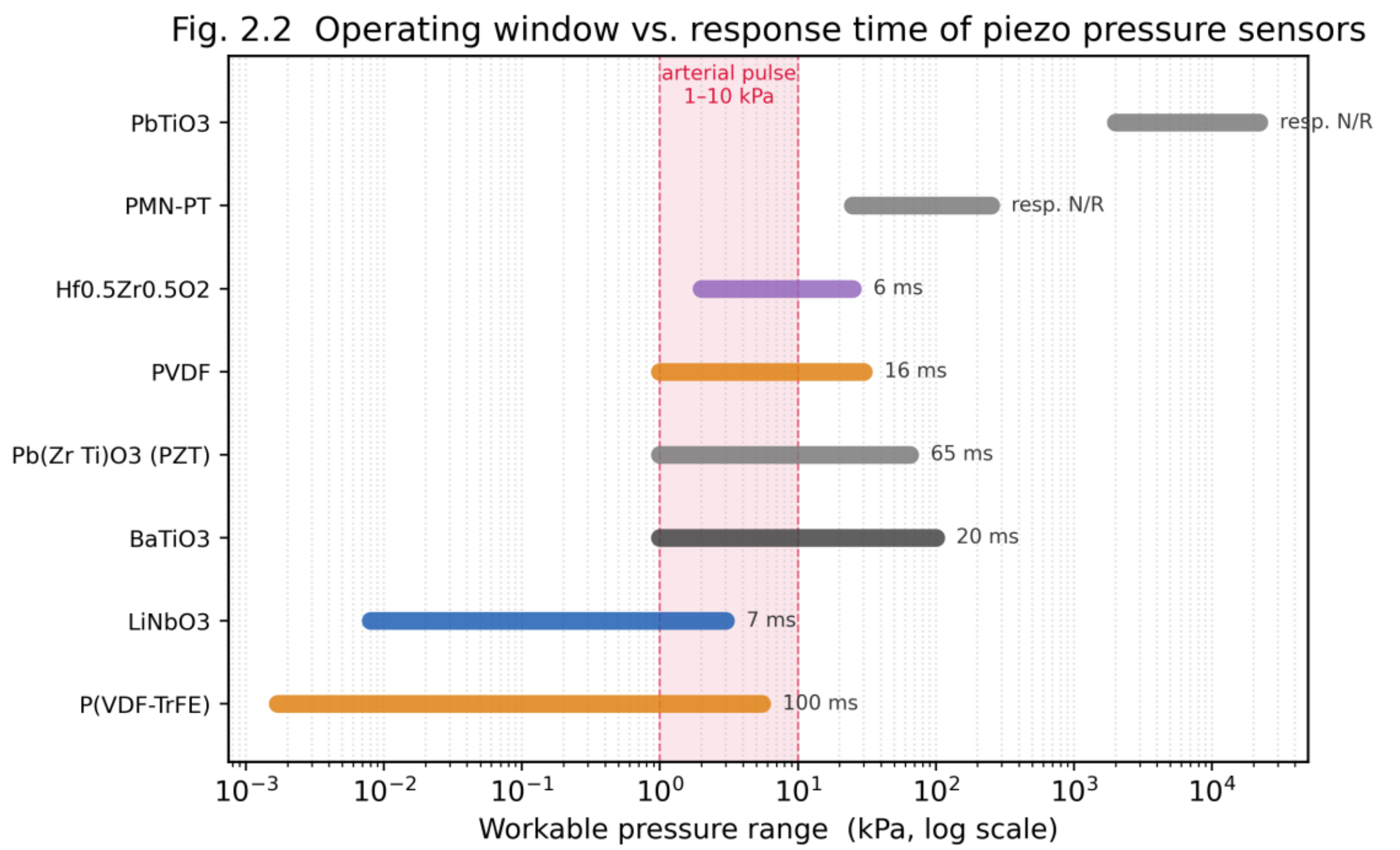

Figure 3. Workable pressure range vs. response time of representative piezoelectric materials; shaded band = 1–10 kPa arterial-pulse window ($BaTiO_3$ and PVDF span it). Data compiled from Deng et al. [10].

## 2.2 Lead-Free Contenders — $BaTiO_3$-based (BZT–BCT) vs. Alkali Niobates (KNN)

The lead-free field has converged on two perovskite families. The $BaTiO_3$-based system, optimized as BZT–BCT, delivers the highest reported lead-free $d_{33}$ (≈620 pC/N at the MPB) and the chemical robustness of a well-behaved oxide, but at the cost of a low Curie temperature, $T_c \approx 93$ °C [18], [19], [29]. The alkali-niobate system, $(K,Na)NbO_3$ (KNN), offers the opposite balance: a high $T_c$ (>400 °C) but a modest base $d_{33}$ (~80–110 pC/N), and a notorious sensitivity to alkali volatilization and poor densification during sintering [27], [30], [31]. A third lead-free family, the bismuth sodium titanate (BNT)-based perovskites, delivers large field-induced strain but is constrained for the present purpose by a depolarization temperature lying well below the Curie point, so a stably poled state is not retained across the whole operating window; we therefore treat it comparatively rather than as a candidate [12], [30].

**The sensitivity–stability trade-off.** The two systems share a single governing trade-off that any honest comparison must foreground: raising $d_{33}$ lowers thermal stability, and vice versa. In KNN this is explicit and quantitative—chemical engineering of the rhombohedral–orthorhombic–tetragonal polymorphic phase transition (PPT) to boost $d_{33}$ to 344–425 pC/N simultaneously drives $T_c$ below 200 °C, while Li/Ta substitutions that preserve $T_c \approx$ 450–510 °C cap $d_{33}$ near 200–280 pC/N; only elaborate template grain growth—the textured-ceramic route established for alkali niobates [32]—reaches $d_{33} \approx$ 700 pC/N, and then with $T_c \approx$ 242 °C [27]. Supplementary Table 1 tabulates this inverse relationship, and Figure 4 maps it across the lead-free families.

**Why** body temperature matters more than it appears. For a skin-worn device the operating point is 310 K (37 °C). Both families sit comfortably below their $T_c$ in absolute terms, so neither depolarizes at body heat. The subtlety—and a point we believe the literature under-states—is proximity to phase boundaries rather than to $T_c$ itself. BZT–BCT owes its giant response to a near-room-temperature confluence of phases; its property surface is consequently steep near 300–310 K, so modest thermal excursions can shift $d_{33}$, permittivity, and linearity (a drift/hysteresis risk); compositions tuned slightly away from the boundary trade peak response for a flatter temperature characteristic, which is precisely the compromise a wearable must navigate [19], [29]. KNN's orthorhombic–tetragonal PPT, often engineered toward room temperature precisely to raise $d_{33}$, creates the same liability [31]. This is exactly why finite-temperature computation at 310 K (Sections 4 and 6), rather than the conventional 0 K calculation, is essential for this application.

**Our choice, stated honestly.** We nonetheless select BZT–BCT as the primary system for three defensible reasons: (i) it reaches soft-PZT-class sensitivity without the templated texturing that the two higher-performing lead-free compositions require, so the response is obtained in a form that can still be comminuted, foamed or dispersed into the compliant architectures of §2.4; (ii) $BaTiO_3$ is the best-established biocompatible piezoelectric oxide [17], [29], satisfying the skin-contact safety mandate of §1.2; and (iii) it avoids the alkali-volatility and densification problems that plague KNN reproducibility. The low $T_c$ is treated not as a disqualifier but as a constraint to be quantified—our central methodological argument.

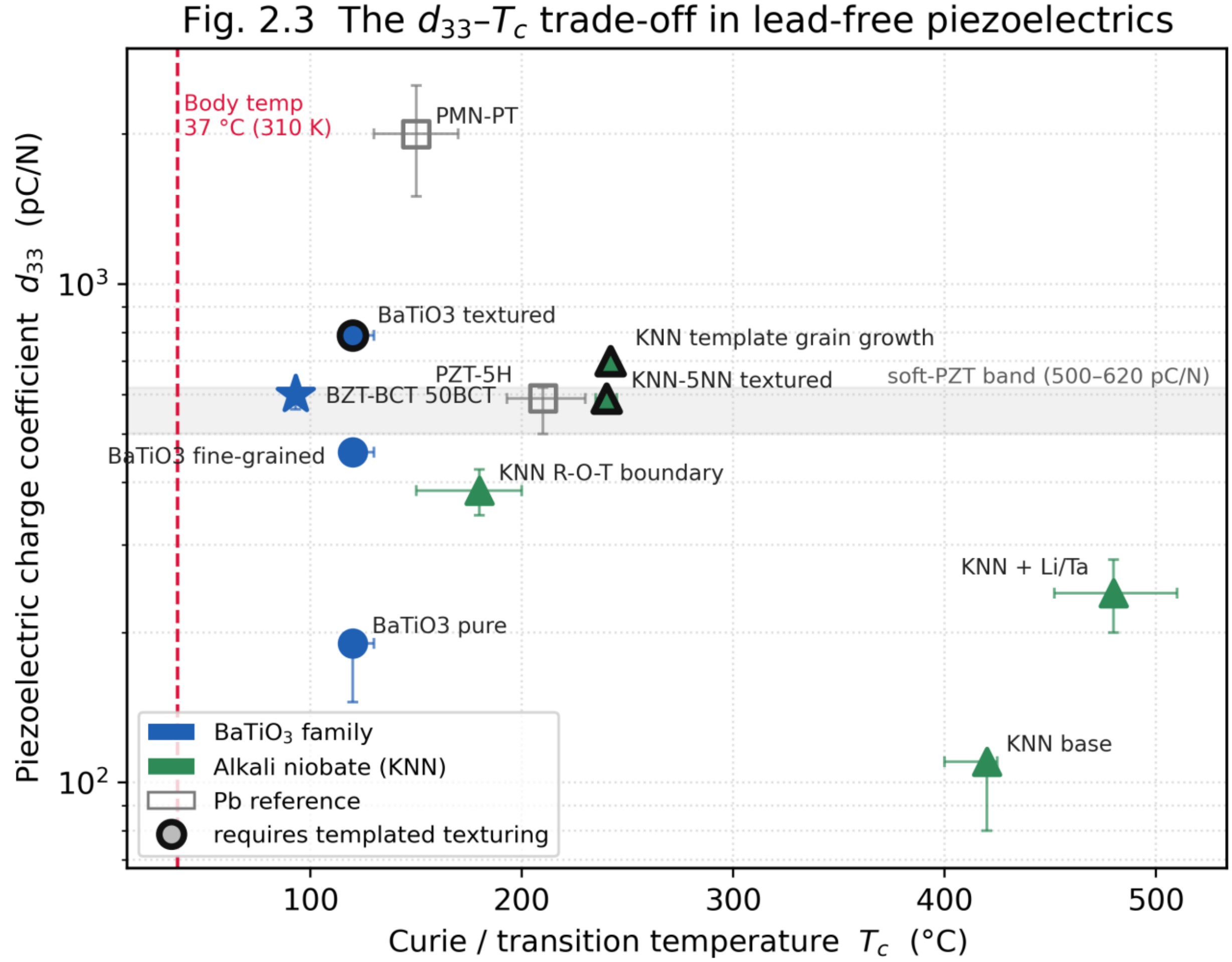

Figure 4. The $d_{33}$–$T_c$ trade-off across lead-free families (log scale). BZT–BCT (star) reaches the soft-PZT band at $T_c \approx 93$ °C; among lead-free ceramics it is matched or exceeded only by textured $BaTiO_3$ and textured / template-grain-grown KNN, and heavy black rings mark exactly those compositions whose $d_{33}$ is attained only through an elaborate templated-texturing route. Shaded horizontal band = soft-PZT range (500–620 pC/N); dashed vertical line = 310 K operating point. Error bars span the literature ranges of Table 1. Data from [18], [19], [27], [29], [30], [32].

## 2.3 The Physics of the Morphotropic Phase Boundary (MPB)

The giant piezoelectricity of BZT–BCT originates in the topology of its phase diagram. Liu and Ren showed that the MPB separating the ferroelectric rhombohedral (R, BZT-rich) and tetragonal (T, BCT-rich) phases emanates from a cubic–rhombohedral–tetragonal (C–R–T) triple point located near $x \approx 0.32$ and $T \approx 57$ °C, and that this triple point is a tricritical point (TCP)—a crossover from first-order to continuous transition at which the energy barrier among the three states vanishes [18]. Subsequent high-resolution synchrotron diffraction has refined this picture: an intermediate orthorhombic phase, isostructural with that of the parent barium titanate, intervenes between the tetragonal and rhombohedral fields, so that the boundary originally assigned as tetragonal–rhombohedral is more properly tetragonal–orthorhombic, and the triple point is better described as a region of phase convergence [33]. The distinction matters for the present argument: it places three, rather than two, nearly degenerate polar phases within a few tens of kelvin of the operating point, which sharpens rather than weakens the finite-temperature case developed in Sections 4 and 6.

**The Landau picture.** The consequence is made precise by a symmetry-adapted Landau expansion of the free energy in the polarization magnitude P and direction n: $F = A(x, T)P2 + B(x, n)P4 + C(x, n)P6$. Tricriticality requires the fourth-order coefficient $B \rightarrow 0$, which flattens the free energy with respect to the polarization magnitude—the mechanism termed polarization extension. Conceptually distinct from this, and equally operative at a morphotropic boundary, is near-degeneracy with respect to the polarization direction—polarization rotation. The two flattening directions should not be conflated, although both are enhanced near a tricritical MPB and both contribute to the measured response [34]. Vanishing polarization anisotropy means there is almost no barrier to rotating the polarization vector between the $\langle 001\rangle$T and $\langle 111\rangle$R states; an external stress or field can therefore reorient the polarization with minimal energy cost, which is the microscopic source of the colossal piezoelectric and dielectric response (Figure 5) [18]. This polarization-rotation mechanism was established from first principles for single-crystal piezoelectrics [35], and is now understood—together with polarization extension—as the general origin of property enhancement wherever the free-energy profile is flattened [34]. In real (slightly tilted) MPBs the anisotropy is small but finite, giving a shallow barrier that is frequently bridged by an intermediate orthorhombic (Amm2) phase [29], [33]—the flattening of the polarization-rotation pathway that experiments and first-principles mapping aim to confirm. High piezoelectricity additionally requires elastic lattice softening, which Liu and Ren note accompanies the low polarization anisotropy [18].

**Link to the flexible device.** For an e-skin sensor this physics is doubly relevant. The barrierless polarization rotation is precisely what allows a feeble 1–10 kPa arterial stress to produce a large polarization change—the device works because the MPB makes the lattice mechanically and electrically "soft." But the very same softness is a near-room-temperature, composition-sensitive phenomenon, so it cannot be assumed to survive intact once the material is diluted into a flexible architecture (§2.4) or warmed to 310 K (§2.2). The MPB therefore defines the intrinsic performance ceiling; the engineering challenge is to approach it in a compliant form.

Figure 2.4   Physics of the morphotropic phase boundary

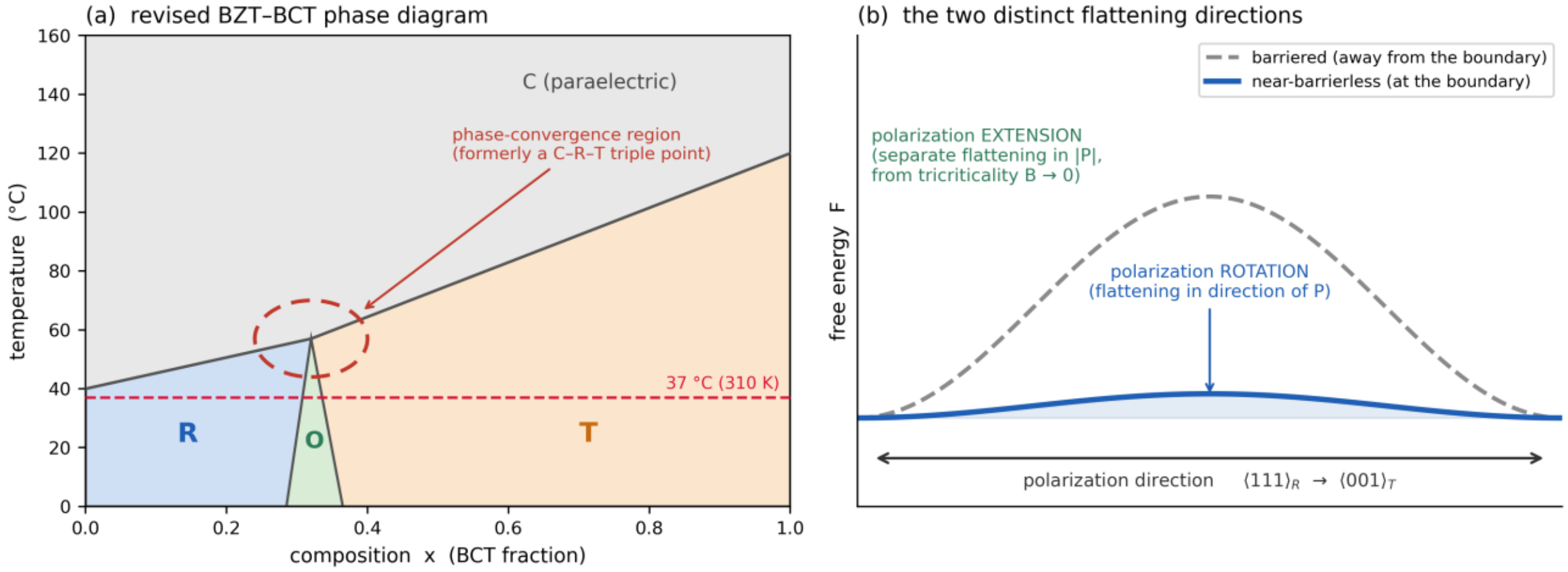

*Figure 5. Physics of the morphotropic phase boundary. (a) Schematic BZT–BCT phase diagram drawn with the revised topology: an intermediate orthorhombic (O) field separates the rhombohedral (R) and tetragonal (T) phases and converges with the cubic (C) paraelectric field in a phase-convergence region near x ≈ 0.32 and ≈ 57 °C, rather than at a single triple point; note that the 37 °C operating line passes immediately below that region. (b) The two distinct flattening directions of the free energy, which should not be conflated: polarization extension, the flattening with respect to |P| associated with tricriticality (B → 0), and polarization rotation, the near-degeneracy with respect to the direction of P that allows ⟨111⟩R → ⟨001⟩T reorientation at small energy cost. Topology after Liu & Ren [18] as revised in ref [33]; mechanisms after refs [34], [35].*

## 2.4 Microstructure & Topology Engineering: Porosity, Nanofibers, Composites

Monolithic BZT–BCT cannot bend; with a Young's modulus of order 100 GPa it is far too stiff and brittle to conform to soft tissue (the ~six-orders-of-magnitude modulus mismatch with skin quantified in §1.3). Four micro-structural strategies are used to confer compliance: (i) ceramic–polymer composites, in which piezoelectric particles, microcubes, or nanowires are dispersed in an elastomer such as PDMS (e.g., NKLN microcube/PDMS harvesters, $d_{33}$ ≈ 460 pC/N for the filler) [36]; (ii) electrospun nanofiber mats, exploited for $BaTiO_3$ pressure sensors with a 1–100 kPa range [10]; (iii) engineered porosity, e.g., $Ba_{0.85}Ca_{0.15}Ti_{0.9}Zr_{0.1}O_3$ foamed with sacrificial carbon-nanotube templates to 3–25 % porosity [20]; and (iv) grain texturing and thin-film growth, which align crystallographic orientation to recover bulk-like coefficients in a conformable form [27], [37]. Figure 6 contrasts the four routes schematically.

**The honest paradox.** These routes succeed mechanically but exact a direct electrical penalty—the point this review insists on stating plainly. Diluting the active ceramic into a low-permittivity polymer, or replacing load-bearing ceramic with void, lowers both the effective $d_{33}$ and the poled volume fraction. The porosity study of the exact BCZT composition is explicit: increasing porosity produces a "dilution effect" that systematically reduces permittivity and the functional response [20]; Figure 7 shows this dilution against the effective-medium bounds. Thus every strategy that buys compliance tends to spend the MPB-enhanced sensitivity that justified choosing BZT–BCT in the first place: the device-level $d_{33}$ of a composite is typically a small fraction of the intrinsic single-crystal value.

**Mitigations (and why computation is still needed).** The trade-off is not perfectly zero-sum, and two subtleties soften it. First, because the voltage coefficient $g_{33}$ scales inversely with permittivity (§2.1), embedding a high-permittivity ceramic in a low-permittivity polymer can raise the effective $g_{33}$ and hence the $V_{oc}$ signal even as $d_{33}$ falls—so a composite can be a better voltage sensor than its filler fraction alone would suggest. Second, texturing is the least lossy compliance route: by aligning grains along the polar axis it recovers a large fraction of the single-crystal coefficient (textured KNN reaches $d_{33} \approx 590$ pC/N), making textured thick-films and the classical connectivity patterns (0–3, 1–3, 2–2) of composite design [38] the most promising compromises [27]. Quantifying where, on this compliance-versus-sensitivity continuum, an optimal architecture lies requires knowing the intrinsic ceiling accurately—returning the problem to first-principles computation and motivating the computational programme of Sections 4–6.

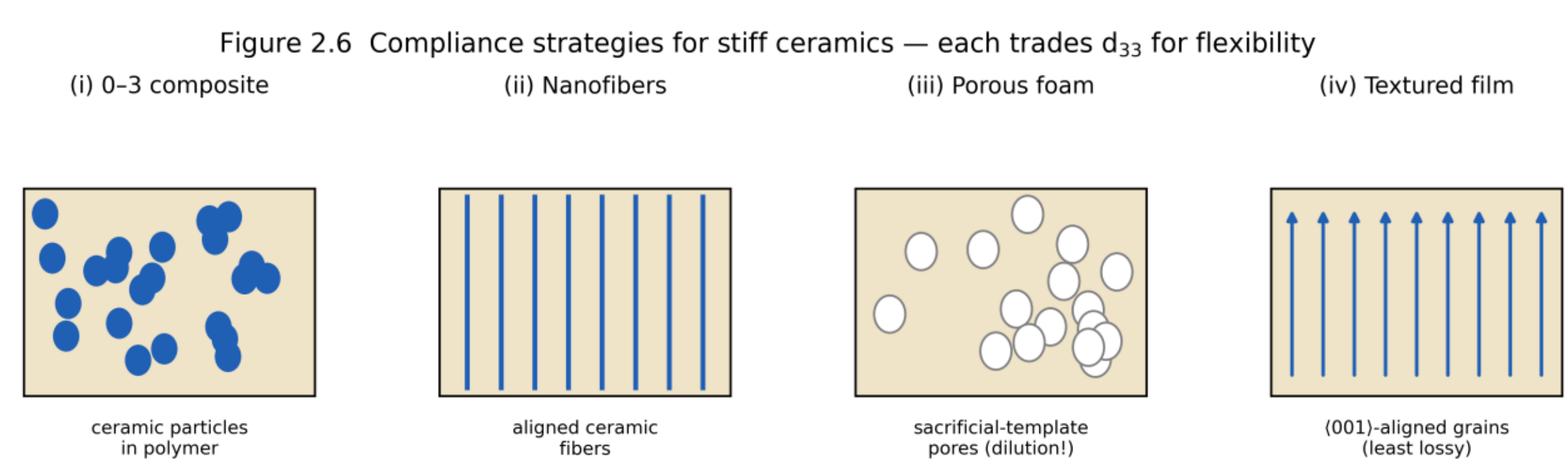

Figure 6. Micro-structural routes to mechanical compliance: (i) 0–3 composite, (ii) nanofibers, (iii) porosity, (iv) texturing. Each lowers stiffness but dilutes the MPB-enhanced $d_{33}$; texturing is the least lossy.

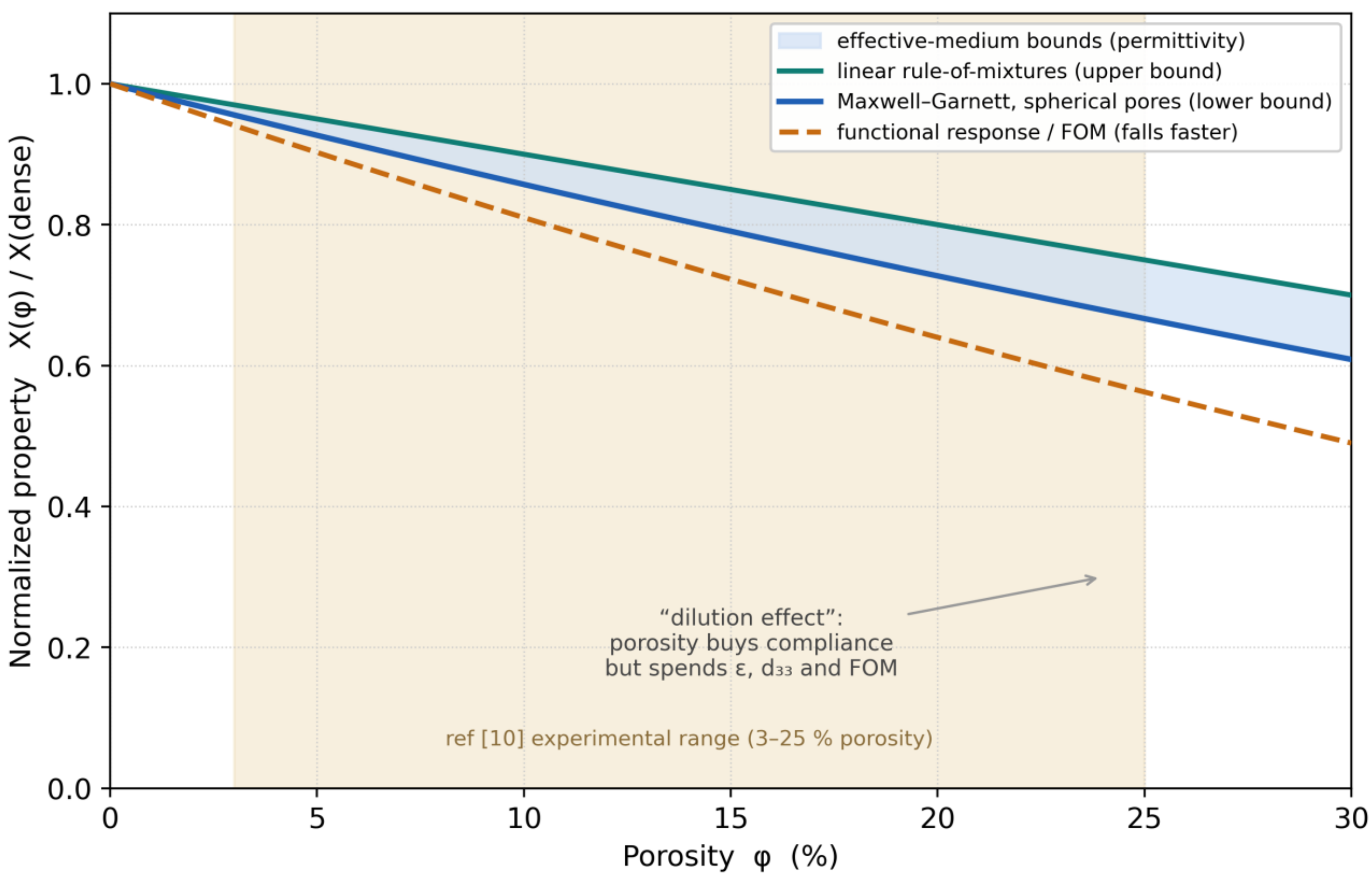

Figure 7. The porosity “dilution” effect on the functional response of porous BCZT. Curves are effective-medium models (linear rule-of-mixtures, upper bound; Maxwell–Garnett spherical-pore model, lower bound) for the normalized permittivity of the porous ceramic; the shaded band marks the 3–25 % porosity range studied experimentally for $Ba_{0.85}Ca_{0.15}Ti_{0.9}Zr_{0.1}O_3$ by Gheorghiu et al. [20], who report the same monotonic reduction of permittivity and functional response with increasing porosity. Data trend and porosity range from ref [20].

## 3. State-of-the-Art: Fabricated Flexible Lead-Free Devices

Sections 1 and 2 argued, from intrinsic material coefficients, that lead-free $BaTiO_3$-based perovskites are the right design choice. This section tests that argument against reality: what has actually been built. We review demonstrated flexible lead-free piezoelectric devices and their measured performance (§3.1), then confront those numbers with the specific requirements of arterial-pulse monitoring to expose the gap this review—and its computational programme—aims to close (§3.2). The distinction from Section 2 is deliberate: there we tabulated intrinsic single-crystal/ceramic coefficients ($d_{33}$, $g_{33}$); here we tabulate device-level metrics (sensitivity in $kPa^{-1}$, output $V_{oc}$, pressure range, response time) of fabricated, flexible objects.

### 3.1 Demonstrated Devices & Measured Performance

**BaTiO**3-based devices. Flexible lead-free piezoelectric devices reported to date cluster into two material families—$BaTiO_3$-based composites and alkali-niobate (KNN) composites—almost always realized as ceramic-in-polymer composites, electrospun nanofiber mats, or nanogenerator (NG) stacks rather than as monolithic ceramic [21], [22]. Figure 8 shows the generic architecture of such a device and the characteristic pulse waveform it produces.

$BaTiO_3$ nanoparticle–carbon-nanotube/PDMS nanocomposite generators reach an open-circuit voltage $V_{oc} \approx 3.2$ V (vs ≈1 V for a bare $BaTiO_3$ thin film) with a ≈50 ms response [39]; the same group later raised the output of lead-free nanocomposite generators by combining $BaTiO_3$ nanowires with nanoparticles as the filler [40], [41]; and bio-inspired 3D-porous BCZT–PDMS composites deliver $V_{oc} \approx 5$ V with a power density of 2.6 µW cm−2 [22]. As pressure sensors, electrospun $BaTiO_3$/polymer composites achieve sensitivities of 0.23–5 kPa−1 with detection limits down to ~0.1 Pa, aligned P(VDF-TrFE) nanofibre arrays supply the polymer benchmark for low-pressure transduction [42], and a polydopamine-coated BTO/PVDF sensor produced 9.3 V with a 61 ms response while resolving a human wrist pulse at 75 beats min−1—a direct cardiovascular demonstration [21].

**KNN-based devices.** KNN composites dominate the lead-free energy-harvesting demonstrations: NKLN microcube/PDMS films reach $V_{oc} \approx 48$ V, textured KNN flexible harvesters ≈ 25 V under cyclic force, and KNN-based triboelectric generators ≈ 70 V from body motion [27], [36], [43]. Most relevant to this review, a P(VDF-TrFE)/KNN/graphene electrospun film was used as a flexible cardiac sound sensor: it generated 1.9 V (38× a bare-PVDF reference), captured heart sounds across 20–320 Hz over 2800 cycles, and—coupled to a classifier—reached ~99 % accuracy in distinguishing normal from abnormal heart sounds [7]. For reference, the most complete demonstrations of conformal, skin-mounted pulse sensing remain lead-based: conformable PZT arrays laminated directly onto the epidermis resolve the arterial pressure waveform with high fidelity [44], and the same device class has been used to map the viscoelastic modulus of soft tissue in a clinical setting [45]. These define the performance target that a lead-free device must meet, and they make the absence of an equivalent lead-free demonstration the more conspicuous.

**What the literature shows—and omits.** Table 2 compiles these fabricated devices, and Figure 9 compares their reported open-circuit outputs by material family. One absence is conspicuous and is the crux of §3.2: although BZT–BCT has the highest lead-free $d_{33}$, it appears almost exclusively as rigid bulk (e.g., a sintered-disc acoustic-emission sensor, 65 dB sensitivity [28]); flexible BCZT composites do nevertheless exist: BCZT–PDMS hybrid nanocomposites and BCZT-loaded PVDF films have both been demonstrated as wearable nanogenerators [46], [47], and a lead-free piezoceramic thin film transferred onto a plastic substrate has been shown to combine high output with comprehensive biocompatibility testing [48]. Critically, however, those devices are reported as energy harvesters and are characterized by open-circuit voltage under finger tapping or shaker excitation, not by a calibrated pressure sensitivity within the 1–10 kPa arterial window; to our knowledge no flexible BZT–BCT arterial-pulse sensor has been demonstrated and characterized as such. The gap is therefore sharper than a simple absence of compliant BZT–BCT: the composition can be made flexible, but its compliant forms have never been evaluated against the pulse-sensing task.

Table 2. Representative fabricated flexible lead-free piezoelectric (and related) devices. Because different studies report different figures of merit, performance is given in one consolidated column exactly as measured in each source, rather than as sparse separate columns. Mode: piezo = piezoelectric; NG = nanogenerator; TENG = triboelectric. The final row underscores that BZT–BCT—the highest-$d_{33}$ lead-free composition—has been realized only as a rigid bulk element, not a flexible device.

| Active material / architecture | Mode | Reported performance (as measured) | Demonstrated application | Ref. |
|---|---|---|---|---|
| **P(VDF-TrFE)/KNN/graphene, electrospun film** | Piezo | 1.9 V (38× bare PVDF); 20–320 Hz; 2800 cycles | Heart-sound diagnosis (~99% acc.) | [7] |
| **Sb–$BaTiO_3$–P(VDF-TrFE) (SBP) nanofiber** | Piezo | 96 mV/kPa; 2 ms; up to 128 kPa; 17.1 V; 2400 cycles | Wearable pressure sensing | [21] |
| **Polydopamine@$BaTiO_3$/PVDF composite** | Piezo | 9.3 V; 61 ms response | Human wrist pulse (75 bpm) | [21] |
| **P(VDF-HFP)/MXene/$BaTiO_3$, gradient fiber** | Piezo | 0.23 $kPa^{-1}$; <1 kPa range; 500 cycles | Low-pressure tactile e-skin | [21] |
| **Electrospun $BaTiO_3$/PVDF composite** | Piezo | 0.23–5 $kPa^{-1}$; LOD ~0.1 Pa; 5000 cycles | Wearable e-skin | [21] |
| **$BaTiO_3$ NP–CNT/PDMS (NCG)** | Piezo NG | 3.2 V; 350 nA; 50 ms | Biomechanical energy harvest | [22] |
| **3D-porous BCZT–PDMS (PCG)** | Piezo NG | 5 V; 0.55 μA/cm²; 2.6 μW/cm² | Self-powered harvester | [22] |
| **KNN (NKLN) microcube/PDMS** | Piezo NG | 48 V; 0.43 μA/cm² @ 2 kgf; ~11% eff. | Implantable power | [36] |
| **Textured KNN (F-PEH)** | Piezo NG | 25 V; 0.4 μA; 5.5 mW/m²; fatigue-tested | Flexible energy harvester | [27] |
| **KNN–Kapton** | TENG | 70 V; 1100 nA; | Body-motion | [43] |

| | | 2.5×2.5 cm active area | harvesting | |
|---|---|---|---|---|
| **BZT–BCT sintered disc (RIGID — not flexible)** | Piezo | $d_{33}$ = 370 pC/N; $g_{33}$ = 11.3 mV·m/N; $k_p$ = 0.58; 65 dB | Acoustic-emission sensor | [28] |

Figure 3.3 Flexible piezoelectric arterial-pulse e-skin and its characteristic output

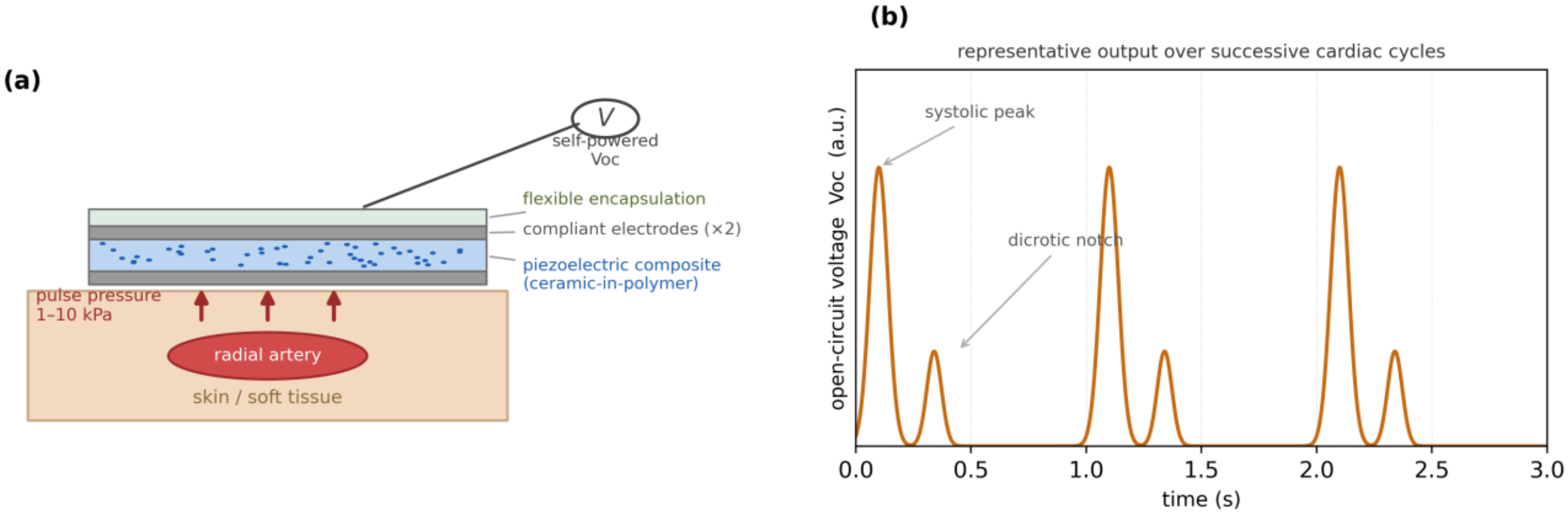

*Figure 8. Schematic of a flexible piezoelectric arterial-pulse e-skin and its characteristic output. (a) A conformal ceramic–polymer piezoelectric film with compliant electrodes and flexible encapsulation, worn over the radial artery, transduces the 1–10 kPa pulse-pressure wave into a self-powered open-circuit voltage. (b) Representative output waveform over successive cardiac cycles, resolving the systolic peak and the dicrotic notch. Representative of the flexible lead-free pulse and heart-sound sensors demonstrated in refs [7], [21].*

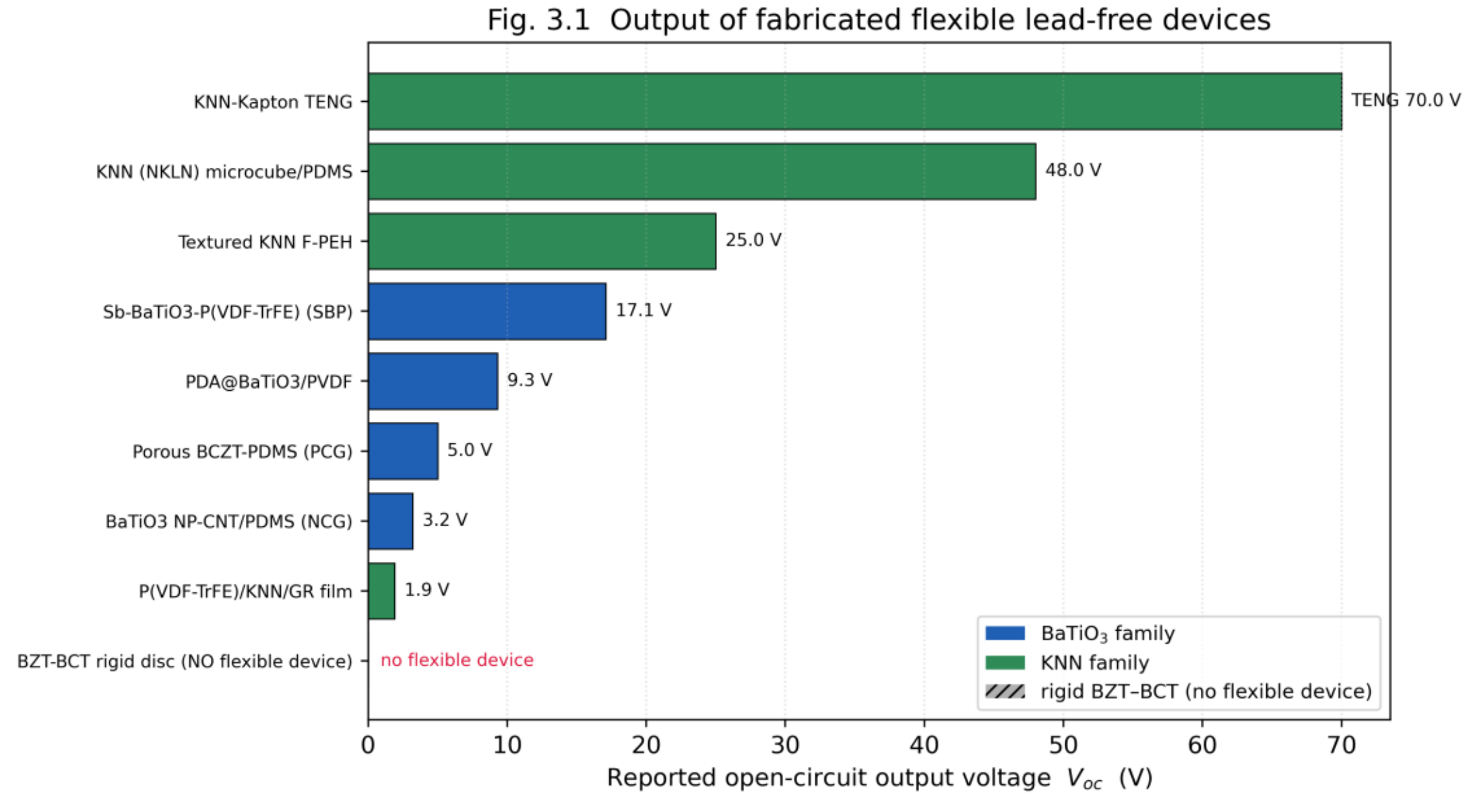

*Figure 9. Reported open-circuit output voltage of fabricated flexible lead-free devices, by material family; the rigid BZT–BCT element (hatched) has no flexible-device counterpart. Data from [7], [22], [27], [28], [36], [43].*

## 3.2 The Gap Between Reported Performance and Clinical Requirements

**Where the field already succeeds.** Reading Table 2 against the clinical mandate of Section 1 exposes a set of specific, addressable gaps rather than a single deficiency. Output voltage is not the problem: self-powered $V_{oc}$ values of 1.9–70 V comfortably exceed what readout electronics require, and lead-free biocompatibility is essentially met by the $BaTiO_3$/KNN/PVDF chemistries. The gaps lie in the precise matching of the transducer to the arterial-pulse task.

**Where the gaps are.** First, pressure-window mismatch: most devices are optimized either for high-pressure tactile sensing (tens to hundreds of kPa) or for sub-kPa touch, whereas the arterial window is the narrow 1–10 kPa band; few devices report calibrated, linear response specifically there. Second, response-time spread: reported values span roughly two orders of magnitude, from a few milliseconds to several hundred milliseconds [21], and the slower devices cannot resolve the dicrotic notch of the pulse wave (which needs ≈20–50 ms). Third, and most important, the high-sensitivity gap: the giant-$d_{33}$ BZT–BCT composition has not been realized as a flexible pulse sensor at all, so the field is leaving its best material on the table. Fourth, two under-reported reliability axes: operating stability at 310 K (body temperature) is rarely characterized in the sources surveyed here—precisely the finite-temperature behaviour Section 2.2 flagged as a BZT–BCT liability—and cyclic durability is typically demonstrated only to 500–5000 cycles, far short of the 105–106 cycles implied by chronic wear. Table 3 sets each clinical requirement against the typical achieved status, and Figure 10 renders the resulting profile across seven axes.

*Table 3. Clinical requirements for an arterial-pulse e-skin versus the typical status of fabricated lead-free flexible devices, with the resulting research gap. The final two rows define the specific openings this review and its computational framework target.*

| **Clinical requirement** | **Target (arterial pulse)** | **Typical lead-free flexible status** | **Gap / action** |
|---|---|---|---|
| **Pressure window** | 1–10 kPa | optimized 0–100+ kPa or <1 kPa; few in-band | Re-tune to 1–10 kPa |
| **In-band sensitivity & linearity** | high, linear | 0.23–5 $kPa^{-1}$ (lead-free) but range-/mode-dependent | Linearize within 1–10 kPa |
| **Response time** | ≈20–50 ms | 2–560 ms (wide spread) | Many too slow for dicrotic notch |
| **Self-powered output** | readable V oc | 1.9–70 V demonstrated | Met |
| **Lead-free + biocompatible** | required | $BaTiO_3$ / KNN / PVDF blends | Met |
| **Giant-$d_{33}$ BZT–BCT in flexible pulse sensor** | desired | absent (only rigid bulk disc) | KEY GAP — target of this review |
| **Stability at 37 °C (310 K)** | characterized & stable | rarely measured at body T | Needs finite-T study (§4, §6) |
| **Cyclic durability** | $10^5$–$10^6$ cycles | 500–5000 typical | Endurance gap |

**Why this gap leads to the rest of the paper.** These gaps motivate the remainder of the article. The high-sensitivity and 310 K-stability gaps cannot be closed by fabrication trial-and-error alone: identifying a BZT–BCT composition and micro-architecture that keeps a usable fraction of the MPB $d_{33}$ while remaining linear and thermally stable in the 1–10 kPa / 37 °C operating envelope is a high-dimensional search problem. That is exactly where the computational programme begins—Section 4 dissects why conventional 0 K DFT cannot tractably perform this search, and Sections 5–6 propose the equivariant-MLIP framework that can.

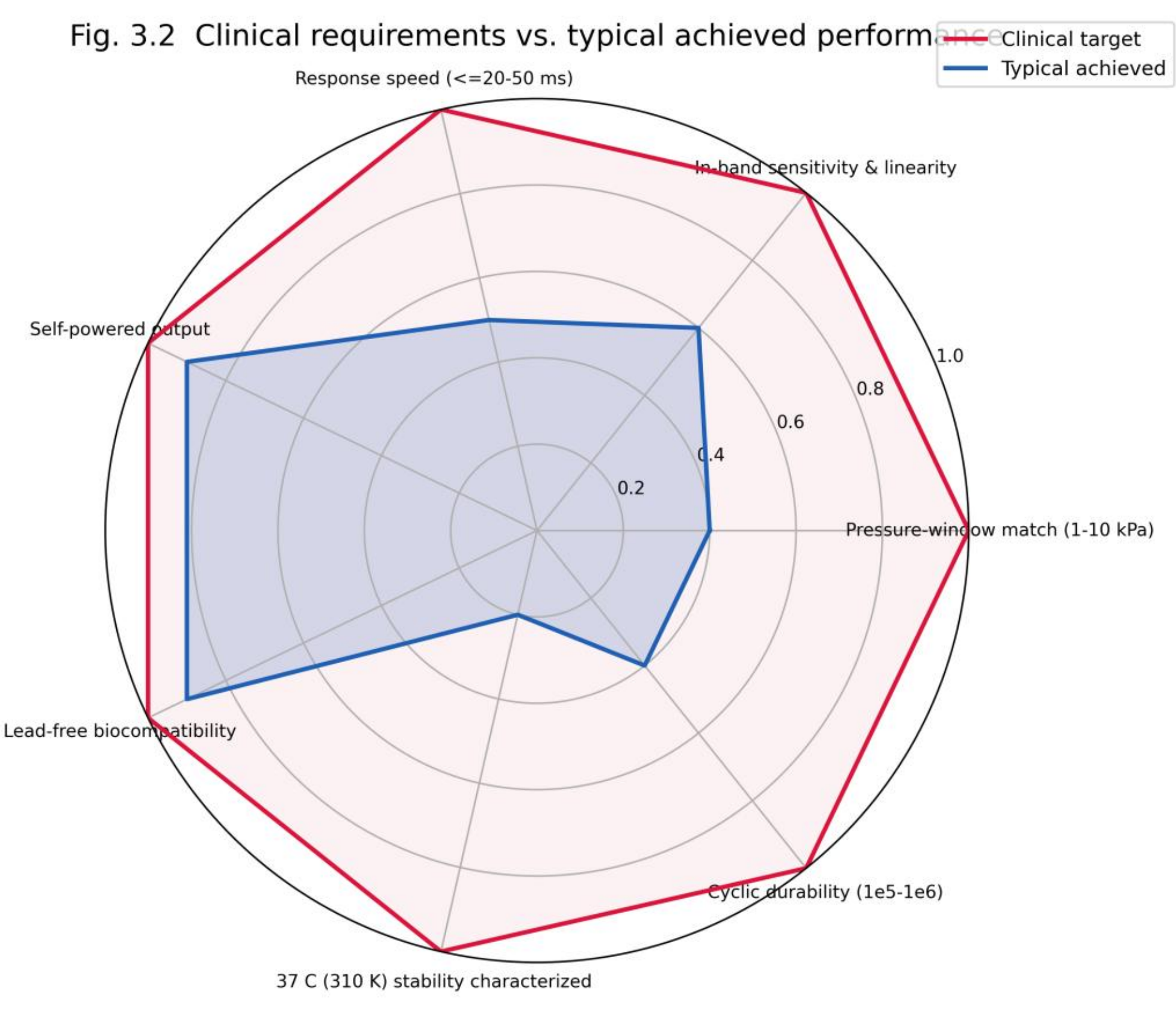

*Figure 10. Clinical requirements (red envelope) vs. typical achieved performance (blue) of lead-free flexible devices across seven axes; the inward dents at pressure-window match, 37 °C stability and cyclic durability define the gaps this review targets. Maturity scores are illustrative.*

## 3.3 Durability, Thermal Stability and Clinical Validation

Two requirements decide whether a demonstrated device becomes a usable one, and both are systematically under-reported in the literature surveyed above. The first is endurance. Cyclic testing of the flexible lead-free devices in Table 2 typically stops between 500 and 5000 cycles [7], [21], [27], whereas continuous wear at a resting heart rate accumulates of order $10^5$ cycles per day, so the reported demonstrations cover a few hours of equivalent service. Ferroelectric ceramics are moreover subject to polarization fatigue and microcrack accumulation under repeated electromechanical loading, a degradation mode characterized in detail for bulk piezoceramic actuators [49]. A pressure sensor operates far below the coercive field, so the switching-driven mechanism does not transfer directly; the point is rather that long-term electromechanical stability cannot be assumed from a short cyclic test, and for a ceramic-in-polymer composite the interface between filler and matrix adds a further, largely uncharacterized, fatigue pathway.

The second is thermal stability at the operating point. Almost none of the devices in Table 2 reports its transfer characteristic at 37 °C, and for BZT–BCT this is the critical omission rather than a minor one: the composition owes its response to a near-room-temperature phase convergence, so its permittivity and $d_{33}$ vary steeply across the narrow band in which the device must work (§2.2, §2.3) [19], [29]. A sensitivity quoted at ambient laboratory temperature therefore does not transfer to the wrist without measurement. Encouragingly, the adjacent problems are tractable: flexible piezoelectric harvesters have operated in vivo over extended periods [50], lead-free thin-film harvesters have passed comprehensive biocompatibility panels [48], and barrier encapsulation of flexible bioelectronics is now a developed engineering discipline with quantified lifetimes [51].

A third gap is one of evidentiary standard rather than of physics, and it separates this literature from clinical acceptance. Resolving a recognizable pulse waveform, which several of the devices above achieve, is not the same as demonstrating measurement validity. Cuffless blood-pressure devices are now governed by explicit protocols—IEEE Std 1708 for wearable cuffless devices [52] and ISO 81060-3 for continuous cuffless measurement against an intra-arterial reference [53]—and the American Heart Association has set out how such devices should be assessed [5]. To our knowledge no lead-free flexible piezoelectric pulse sensor has been evaluated against any of these. Closing that gap is a device-validation task rather than a materials one, but it defines the standard of evidence that the materials programme of the following sections must ultimately serve.

## 4. The Computational Bottleneck in Solid-Solution Perovskites

Sections 1–3 established a clear materials target: a flexible, lead-free, BZT–BCT-based transducer that keeps a usable fraction of the morphotropic-phase-boundary (MPB) $d_{33}$ while remaining linear and stable in the 1–10 kPa / 37 °C operating envelope. Identifying such a composition and architecture is, at heart, a first-principles search problem—and it is here that conventional density functional theory (DFT) hits three compounding walls: the combinatorial explosion of atomic configurations (§4.1), the systematic band-gap error of affordable functionals that corrupts any leakage/insulation estimate (§4.2), and the 0 K nature of standard DFT versus the 310 K physiology of the device (§4.3). Crucially, all three walls point to the same escape route—a cheap, accurate surrogate for DFT energies and forces—motivating the machine-learning framework of Sections 5–6.

### 4.1 The Configurational Entropy Challenge

BZT–BCT is a substitutional solid solution: at the x = 0.50 MPB the supercell stoichiometry is $Ba_{0.85}Ca_{0.15}Ti_{0.90}Zr_{0.10}O_3$, with Ca randomly substituting 15 % of the A-site Ba and Zr substituting 10 % of the B-site Ti. The number of distinct ways to arrange those dopants on the perovskite lattice grows combinatorially with cell size. For an N-formula-unit cell the count is C(N, nCa) × C(N, nZr), and Supplementary Table 2 shows how violently this explodes.

**Why this is the primary wall.** The implication is stark. Even the modest 3×3×3 cell needed to host the correct composition presents ~5 × 107 raw arrangements—~4 × 104 after symmetry—against a realistic DFT budget of perhaps a few hundred relaxations. The comparison is worth making concrete. The community-wide high-throughput repositories that underpin modern materials informatics—the Materials Project [54], the Open Quantum Materials Database [55] and Alexandria [56]—together hold of order $10^5$–$10^6$ relaxed structures, accumulated over more than a decade of sustained, distributed effort. That entire collective output is still some thirteen orders of magnitude smaller than the number of dopant arrangements available to a single 4×4×4 cell of one composition. The configurational problem is therefore not one that more computing time, or a larger community database, can be expected to solve. Larger cells, required to represent the disorder faithfully and to suppress finite-size artefacts in the elastic and piezoelectric tensors, are hopeless by brute force (1019–1038 arrangements). The associated configurational entropy, $S_{\text{config}} = k_B \ln \Omega$, is correspondingly large and physically real—it is part of what stabilizes the disordered solid solution. The conventional workaround is to abandon the ensemble and compute a single representative structure—a special quasirandom structure (SQS), constructed so that its short-range correlation functions match those of the ideal random alloy [57], [58]—or a handful of hand-built cells; the alternative of parametrizing the configurational energy through a cluster expansion [59], [60] removes that bias only at the cost of a DFT training set of its own. That is tractable but introduces configurational bias: properties are reported for one arrangement that may or may not represent the true low-energy ensemble. Figure 11 visualizes the explosion against the DFT-feasible band and frames the core problem this review addresses: how to survey, rather than sample, the configurational space.

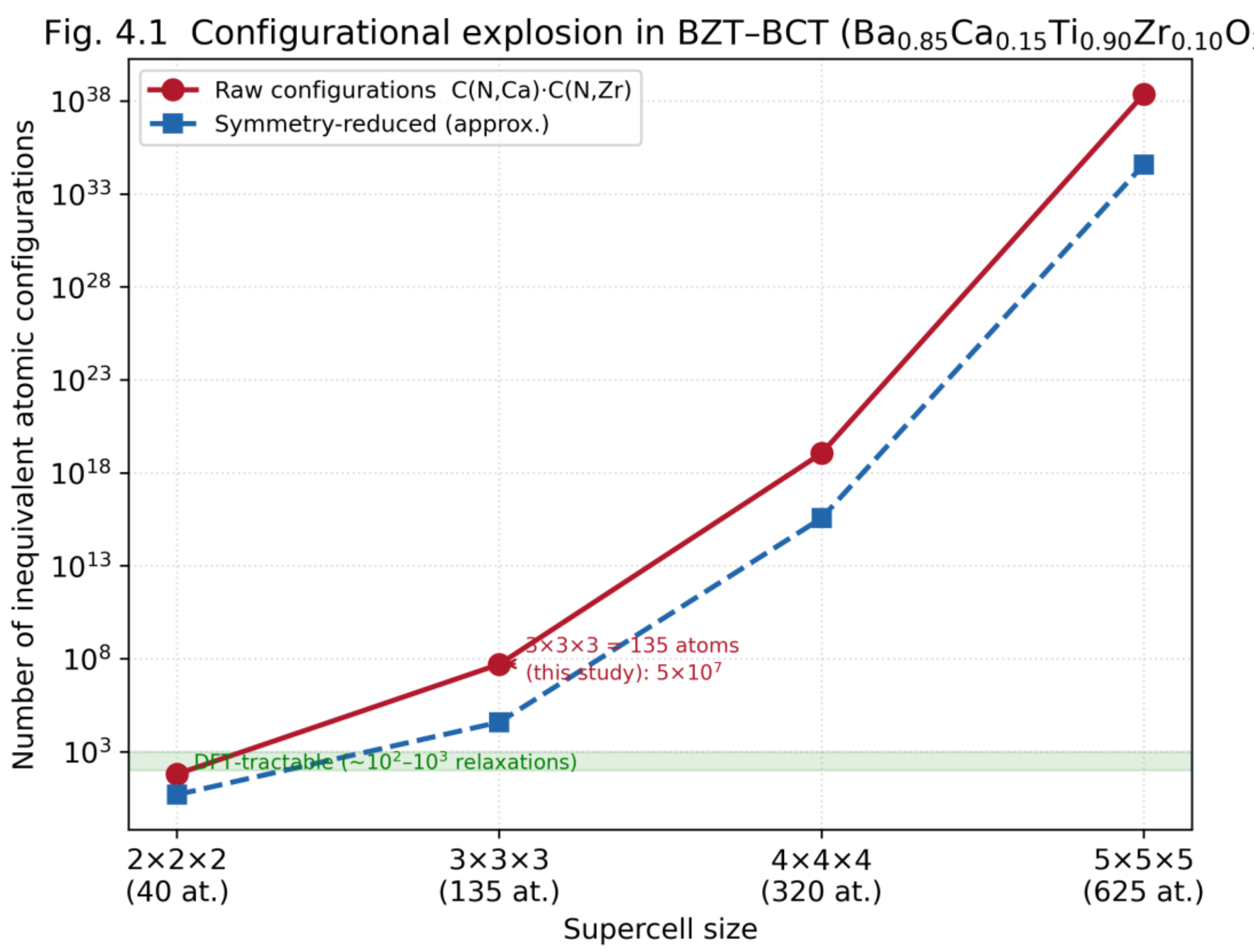

*Figure 11. Combinatorial explosion of dopant arrangements in BZT–BCT vs. supercell size (log scale), against the DFT-tractable band (~$10^2$–$10^3$). Counts computed exactly.*

## 4.2 Bandgap Underestimation, Leakage Current & Insulation Safety

The second wall concerns accuracy, not cost. Affordable semilocal functionals (LDA, PBE [61], PBEsol [62]) systematically underestimate the electronic band gap $E_g$ by 30–50 % because of the self-interaction error inherent in the approximate exchange [63]. For $BaTiO_3$ the experimental gap is ≈ 3.2 eV, whereas PBE returns ≈ 1.7–1.8 eV (Supplementary Table 3 and Figure 12). The meta-GGA r²SCAN [64], a numerically efficient revision of SCAN [65], improves this to ≈ 2.0–2.4 eV; only the screened hybrid HSE06 recovers the experimental value (≈ 3.0–3.3 eV)—at roughly fifty- to a hundred-fold the cost.

**What the band gap actually governs (and what it does not).** It is essential to state precisely why the gap matters here, because the materials-for-sensors literature often over-claims. The band gap does NOT directly determine "patient safety"—chronic biocompatibility and the prevention of any electrical hazard to the wearer are functions of device encapsulation, electrode chemistry and current-limiting, not of the active layer's intrinsic $E_g$. What the gap DOES govern is the insulation quality of the piezoelectric layer: its leakage-current density and its resistance to charge bleed-off. Intrinsic carrier leakage scales as roughly $\exp(-E_g/2kBT)$ [66], and tunnelling/Schottky-limited leakage scales with the gap and the associated band offsets; an $E_g$ underestimated by ~1.5 eV therefore over-predicts leakage by many orders of magnitude and can mislabel a perfectly good insulator as lossy. In practice, leakage in real $BaTiO_3$-based ceramics is dominated by extrinsic defects and grain boundaries (e.g., oxygen vacancies) rather than by intrinsic carriers [67]; the band gap nonetheless sets the intrinsic-leakage floor and the energetics of those defect levels, so an accurate $E_g$ remains a prerequisite for any quantitative leakage estimate. For a self-powered sensor this is not academic: leakage drains the piezo-generated charge before it can be read, degrading the low-pressure (1–10 kPa) signal-to-noise ratio.

**The practical consequence.** The resolution is not to run HSE06 on everything—that is unaffordable for the configurational search of §4.1—but to deploy a functional hierarchy: cheap PBEsol/r²SCAN for geometry and energetics across many candidates, with HSE06 single-point gaps reserved for the final down-selected structures whose insulation must be quantified. This tiered strategy is formalized in Section 6.2.

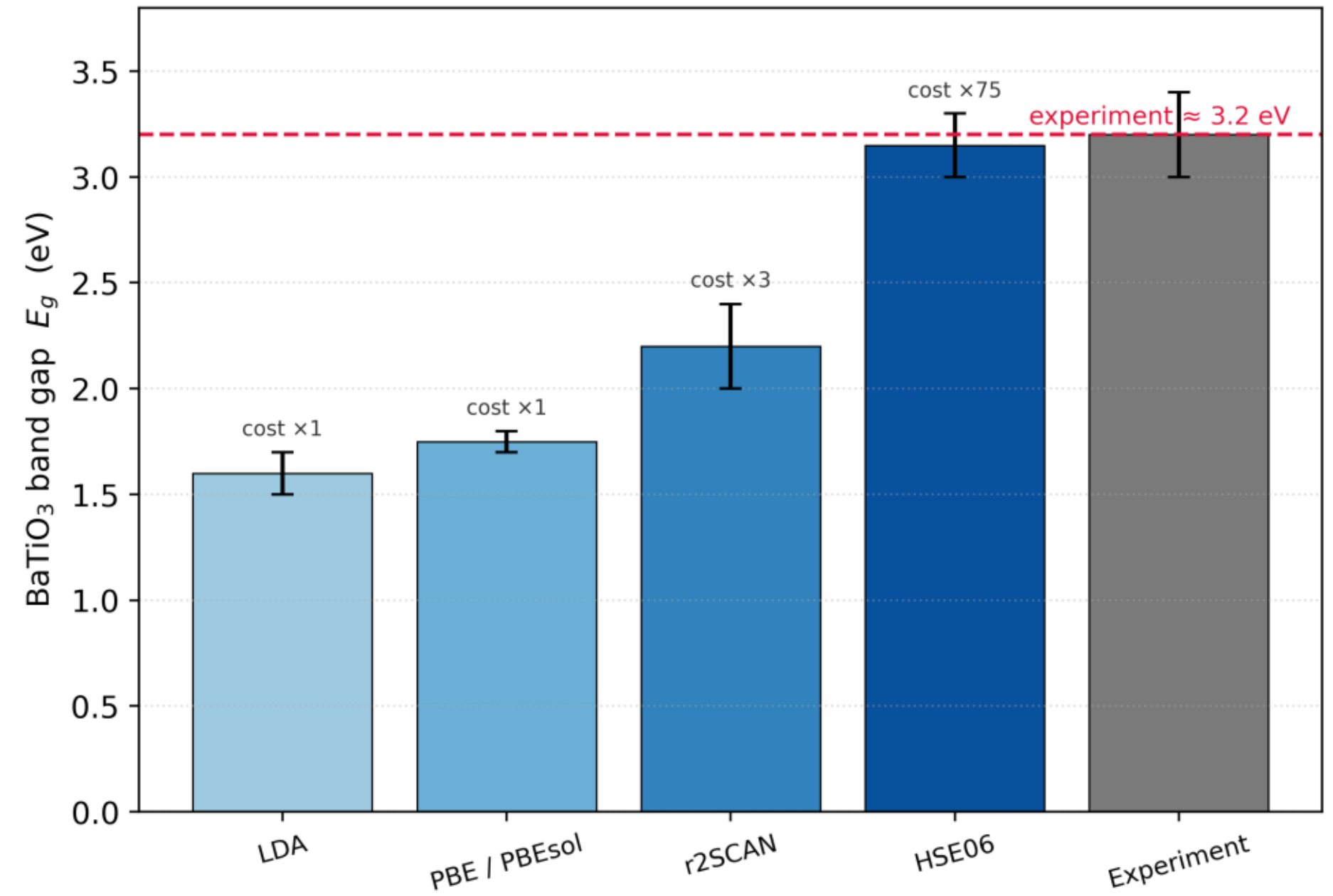

Figure 12. $BaTiO_3$ band gap by functional: PBE/PBEsol underestimate by ~44 %; only HSE06 recovers experiment (~3.2 eV) at ~50–100× cost. Representative literature values [61]–[65], [68], [69].

### 4.3 The 0 K Limitation vs. Physiological Temperature (310 K)

The third wall is temperature. Standard DFT evaluates a static lattice at 0 K: no thermal expansion, no phonon entropy, no thermal population of competing phases. The device, however, operates at 310 K (37 °C), and BZT–BCT is exactly the system for which this matters most, because its giant response originates in a near-room-temperature confluence of phases (§2.3) with a Curie point of only ≈ 93 °C (366 K).

**The energy-scale argument.** The decisive number is the comparison of energy scales. At the MPB the rhombohedral, tetragonal and (intermediate) orthorhombic phases are near-degenerate, separated by only ~2–10 meV per formula unit between the ferroelectric R/O/T phases [70], [71]. Thermal energy at body temperature is $k_B \times 310$ K = 26.7 meV — several times (up to roughly an order of magnitude) larger than those differences (Figure 13). A 0 K "ground-state" ranking is therefore physically meaningless at 310 K: the phases are thermally accessible and coexist/fluctuate, which is precisely the microscopic origin of the easily-rotated polarization and, simultaneously, of its temperature sensitivity. Reporting a single 0 K ground state would both misidentify the operating phase and miss the drift behaviour flagged in §2.2. Table 4 sets out, quantity by quantity, what a 0 K calculation yields against what 310 K operation requires, and which method bridges each gap.

*Table 4. What 0 K DFT yields versus what 310 K operation requires, and the method that bridges each gap. The recurring need for affordable finite-temperature sampling again points to a machine-learning surrogate (Section 5).*

| Quantity | 0 K DFT yields | Needed at 310 K | Bridge method |
|---|---|---|---|
| **Phase stability (R/T/O/C)** | static ΔE ranking (~2–10 meV at MPB) | Gibbs free energy F(T) incl. phonon + config. entropy | QHA / MLIP-MD / SSCHA |
| **Lattice parameters** | 0 K equilibrium volume | thermally expanded lattice at 310 K | QHA thermal expansion |
| ***$d_{33}$* / permittivity** | static (clamped- + relaxed-ion) | T-dependent, near-MPB softening | finite-T MLIP-MD / effective Hamiltonian |
| **Band gap** | bare 0 K electronic gap | electron–phonon-renormalized gap | Allen–Heine–Williams / special displacement [72] |

**Why QHA is only a first step—and where this leads.** The quasi-harmonic approximation (QHA, via Phonopy [73]) is the entry point to finite-temperature free energies, but it assumes each phonon frequency depends only on volume and cannot describe the imaginary soft modes of the high-symmetry reference phase [74] or the strong anharmonicity near the transition—exactly the regime that governs BZT–BCT. Capturing this requires explicit anharmonic sampling (MLIP-driven molecular dynamics, TDEP [75], or the SSCHA [76]), each of which demands thousands to millions of energy/force evaluations. Brute-force DFT cannot supply them. Thus the three walls converge: the configurational search (§4.1), the accuracy/cost trade (§4.2), and the finite-temperature requirement (§4.3) are all solved by the same instrument—an equivariant machine-learning interatomic potential that delivers near-DFT energies and forces at a tiny fraction of the cost. That instrument is the subject of Section 5.

Figure 4.3 Why 0 K ranking fails at body temperature

Figure 13. Energy-scale argument: thermal energy kBT at 310 K (26.7 meV) exceeds the ~2–10 meV inter-phase (R/O/T) energy differences [70], [71], so a 0 K ground-state ranking is not meaningful at body temperature.

## 5. The Paradigm Shift: Equivariant Graph Neural Networks (GNNs) and Machine-Learning Interatomic Potentials

The surrogate that Section 4 showed to be necessary—one reproducing DFT energies and forces at negligible cost—is a machine-learning interatomic potential (MLIP); the field has since shifted from bespoke, system-specific potentials to universal "foundation" models pre-trained on Materials-Project-scale data. This section sets out their foundations (§5.1), delimits precisely what they can and cannot predict—the single most over-claimed point in the sensor literature (§5.2)—and shows how they turn the intractable configurational search into a routine one (§5.3) while changing how candidate structures are selected in the first place (§5.4).

### 5.1 Foundations of Equivariant Machine-Learning Interatomic Potentials

An MLIP expresses the total energy as a sum of atomic contributions, each a learned function of the local atomic environment fitted to a database of DFT energies and forces [77]. The approach matured through a generation of bespoke, system-specific potentials—Gaussian approximation potentials [78], moment-tensor potentials [79] and deep-potential models [80]—each fitted to a single chemistry at a time. The decisive modern advance is symmetry: leading potentials such as NequIP [81], Allegro [82] and MACE [83] use E(3)-equivariant message passing, in which the internal features transform correctly under rotation, translation and inversion. Because the physical symmetry is built into the architecture rather than learned from data, these models are markedly more data-efficient and accurate—the energy is invariant and the forces and stresses are equivariant by construction.

Trained on large, periodic-table-spanning DFT datasets—chiefly the high-throughput repositories introduced in §4.1 [54]–[56]—such architectures yield universal potentials—M3GNet [84], the charge-informed CHGNet [85], and the equivariant MACE-MP-0 [83]—that treat an arbitrary composition with a single model. For perovskite oxides specifically, the UniPero deep-potential model reproduces the experimental temperature-driven ferroelectric transition sequences of solid solutions containing up to six cations, including relaxor PIN–PMN–PT [86]. The landscape has broadened rapidly since: graph networks trained at scale have been used to expand the catalogue of predicted stable inorganic crystals by an order of magnitude [87], and successive foundation models have been released with progressively larger and more chemically diverse training sets, alongside architectural work aimed at the parallel scalability that large-supercell molecular dynamics demands [88]. Typical accuracies are a few meV/atom in energy and tens of meV/Å in force relative to DFT, at three to six orders of magnitude lower cost [83], [85], [89].

## 5.2 What MLIPs Can and Cannot Predict — The Division of Labor

**What MLIPs deliver.** A trained MLIP returns the potential energy, the atomic forces and the stress tensor (CHGNet additionally predicts site magnetic moments and charge states) [85]. From these primary quantities follow structural relaxation, elastic constants, phonons, thermodynamic stability, thermal expansion and—through molecular dynamics—finite-temperature free energies and phase populations.

**What they do not.** They do not, however, return the electronic and polarization observables that define a piezoelectric sensor: the band gap $E_g$, the spontaneous polarization $P_s$, the Born effective charges $Z^*$, the dielectric tensor ε, and therefore the piezoelectric coefficient $d_{33}$. These derive from the electronic wavefunction and the Berry-phase theory of polarization, and remain the province of DFT/DFPT or of a separate, purpose-trained tensorial-property network [90]; the data-driven study of $BaTiO_3$'s dielectric response, for instance, required a dedicated polarization model alongside the energy potential [91]. Claims that a universal MLIP can "screen for giant piezoelectricity in seconds" conflate energy prediction with property prediction and are, as stated, incorrect. Table 5 summarizes the native outputs of the leading universal potentials and, equally important, what they do not provide.

Table 5. Representative universal MLIPs and the scope of their predictions. All return energies, forces and stresses (hence structural, thermodynamic and finite-temperature quantities); none returns the electronic/polarization observables (E_g, P_s, $d_{33}$, ε) that define a piezoelectric sensor — those require DFT/DFPT or a dedicated tensor-property model (§5.2).

| Model | Architecture | Native outputs | Does NOT provide | Ref. |
|---|---|---|---|---|
| **M3GNet** | graph NN (invariant) | E, F, σ | E_g, P_s, $d_{33}$, ε | [84] |
| **CHGNet** | graph NN, charge-informed | E, F, σ, magmom/charge | E_g, P_s, $d_{33}$, ε | [85] |
| **MACE-MP-0** | E(3)-equivariant MPNN | E, F, σ | E_g, P_s, $d_{33}$, ε | [83] |
| **NequIP / Allegro** | E(3)-equivariant | E, F, σ | E_g, P_s, $d_{33}$, ε | [81] |
| **UniPero (perovskites)** | deep potential + attention | E, F, σ; finite-T MD | E_g, P_s, $d_{33}$, ε | [86] |

The correct architecture is therefore a division of labour (Figure 14, formalized in Section 6): the MLIP performs the massive configurational and finite-temperature sampling, while DFT/DFPT—or a dedicated property network—computes $E_g$, $P_s$ and $d_{33}$ only for the small set of survivors.

Figure 5.1 Division of labour between MLIP screening and DFT property evaluation

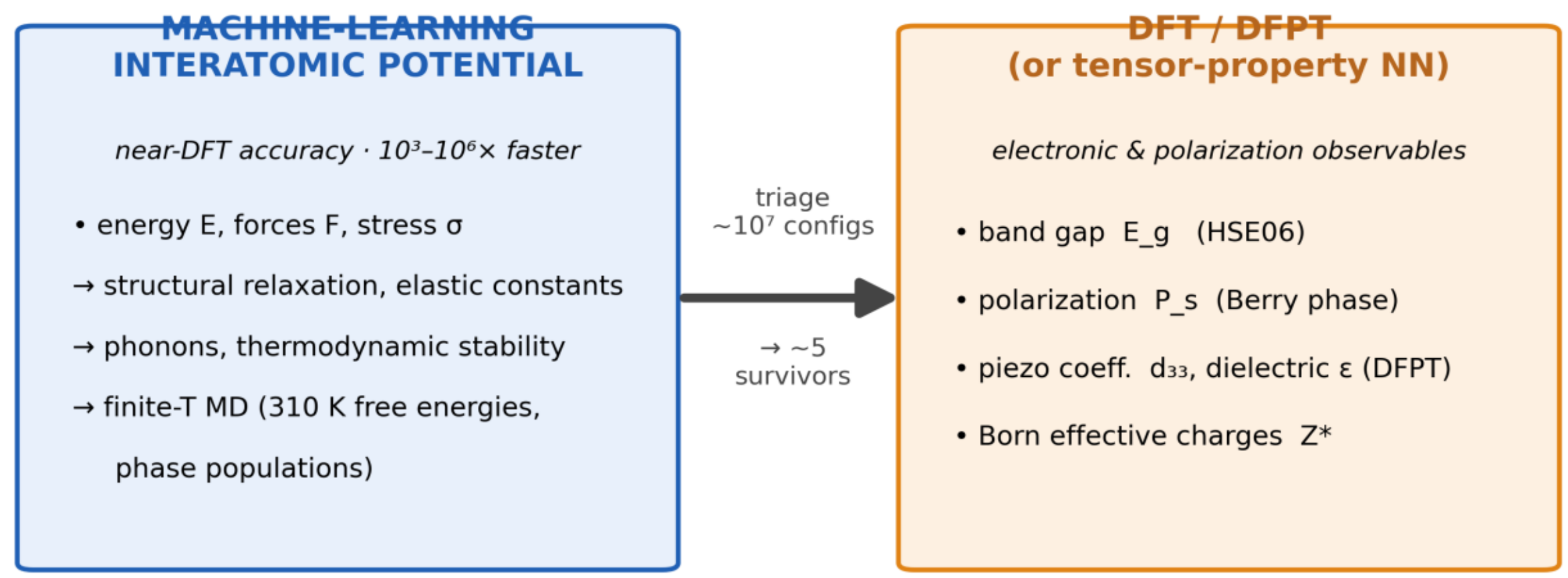

*MLIPs SAMPLE the configurational & thermal space; DFT/DFPT COMPUTES the sensor properties*

*Figure 14. The division of labour between MLIP screening and DFT property evaluation: MLIPs sample the configurational and thermal space; DFT/DFPT (or a tensor-property network) computes the sensor observables.*

## 5.3 Bridging the Micro–Macro Gap: Accelerating Configurational Screening

The configurational space that defeats DFT (§4.1)—from ~5×107 arrangements in the 3×3×3 cell to ~1038 in a 5×5×5 cell—is routine for an MLIP, which evaluates of order 104–106 configurations per CPU-hour on current hardware. The full set of symmetry-inequivalent orderings at a target composition can thus be relaxed and energy-ranked in hours, rather than being reduced to a single hand-built structure. Finite-temperature behaviour then follows from MLIP-driven molecular dynamics (or MLIP-parametrized TDEP/SSCHA), delivering the 310 K free energies and phase populations that the quasi-harmonic approximation cannot (§4.3). Whether that step is trustworthy is now an empirical question rather than an assumption: a systematic benchmark of universal potentials against density-functional reference phonons finds that the better models reproduce harmonic phonon properties with useful accuracy, while the spread between models remains substantial—so the choice of potential, and its validation on the target chemistry, matters as much as the decision to use one at all [92]. UniPero's reproduction of the experimental transition sequences of multi-cation solid solutions is direct evidence that this works for the BZT–BCT class [86].

**Maturity and caveat.** Universal MLIPs already drive high-throughput stability screening across materials databases, and their performance is now tracked on standardized benchmarks such as Matbench Discovery [89], so the approach is established rather than speculative. Its principal limitation is domain of validity: doped solid-solution perovskites such as BZT–BCT are under-represented in the public training sets, so quantitative energy ranking requires fine-tuning or active learning on a modest, chemistry-specific DFT dataset—an explicit step in the framework of Section 6.

### 5.4 Reducing Human Structural-Selection Bias — and the MLIP's Own Biases

Beyond speed, MLIPs change the epistemics of the calculation. The conventional workflow requires a human to choose one Special Quasirandom Structure or a few "reasonable" orderings, so every reported property is conditioned on that choice—the configurational bias identified in §4.1. An MLIP can instead score the entire inequivalent ensemble, or a Boltzmann-weighted sample of it, so that the reported properties reflect the statistical ground state rather than a hand-picked microstate. This is the shift from biased sampling to systematic survey on which the thesis of this review rests.

**An honest counter.** MLIPs are not bias-free, and the limitation is now documented quantitatively rather than merely suspected. They inherit the biases of their training data—the chemistries included and the DFT functional used to generate the labels—and can extrapolate unreliably outside that domain. More specifically, the three universal potentials on which a framework of the kind proposed here would most naturally rest—M3GNet, CHGNet and MACE-MP-0—share a systematic softening of the potential-energy surface: because their pre-training sets are dominated by near-equilibrium configurations, they underpredict the curvature of the energy landscape and consequently underestimate energies and forces in precisely the regimes that matter here, among them solid-solution energetics and phonon vibrational modes [93]. For the present application this is not a peripheral caveat but a direct threat to both load-bearing tasks of Section 6—the energy ranking of dopant configurations and the finite-temperature lattice dynamics—and it is the strongest technical argument for the chemistry-specific fine-tuning and active learning prescribed in §6.5, which has been shown to mitigate the softening [93]. Fine-tuning carries a hazard of its own, however: adapting a general model to a narrow chemistry can erode the performance it had elsewhere, and frameworks that explicitly guard against this forgetting are an active line of work [94]. The practical implication for the present programme is that the fine-tuned potential must be re-validated, not assumed to inherit the generality of the model it was derived from. A systematic survey is therefore only as trustworthy as the potential's validity range, which is why uncertainty quantification and targeted DFT spot-checks are indispensable companions to any MLIP screen (Section 6).

## 6. The Proposed High-Throughput Quantum–AI Framework

Sections 4 and 5 diagnosed the problem and named the instrument; this section assembles them into a concrete, staged workflow. The guiding principle is the division of labour established in §5.2, deployed as a screening funnel (Figure 15): a fine-tuned machine-learning interatomic potential (MLIP) performs the massive configurational and finite-temperature sampling that is intractable for density functional theory (DFT), while quantum-mechanical calculations—reserved for a progressively smaller set of survivors—supply the energetic, electronic and tensorial observables the potential cannot. Each of the three walls of §4 is met by a distinct tier: configurational sampling (§6.1) confronts the combinatorial explosion, the multi-tier DFT hierarchy (§6.2) confronts the accuracy–cost trade-off, and finite-temperature lattice dynamics (§6.3) confronts the 0 K limitation; the sensor observables are then computed on the survivors (§6.4) and the whole pipeline is anchored to data provenance, benchmarking and uncertainty quantification (§6.5); the resulting staged protocol is set out tier by tier in Table 6. We stress at the outset that this is a proposed framework: its individual components are each established in the broader materials-simulation literature, but their integrated application to flexible lead-free BZT–BCT has not yet been carried out, and we flag throughout where the approach remains unproven.

### 6.1 Configurational Sampling Strategy: SQS vs. Cluster Expansion vs. MLIP Sampling

The first decision is how to represent the substitutional disorder of $Ba_{0.85}Ca_{0.15}Ti_{0.90}Zr_{0.10}O_3$ (§4.1). Three strategies exist, and—because they answer different physical questions—conflating them is the source of a common inconsistency in the literature. (i) A special quasirandom structure (SQS) is the single periodic supercell whose short-range correlation functions best reproduce those of the ideal random alloy [57]; modern implementations generate it by Monte-Carlo optimization over both the site occupations and the supercell shape [58]. An SQS is inexpensive and well suited to the average properties of the maximally disordered state, but it is one structure: it neither enumerates the configurational ensemble nor exposes any tendency to order. (ii) A cluster expansion (CE) parametrizes the configurational energy as a generalized Ising Hamiltonian fitted to a set of DFT-computed orderings [59], [60]; coupled to Monte-Carlo sampling it yields the configurational thermodynamics—short-range order, mixing energetics and order–disorder transition temperatures—that an SQS cannot. Its price is the DFT training set and the care needed to converge the cluster basis. (iii) Direct MLIP sampling relaxes and energy-ranks the full set of symmetry-inequivalent orderings (or a Boltzmann-weighted subset of it) at near-DFT accuracy, processing $10^4$–$10^6$ configurations per CPU-hour (§5.3).

These strategies are complementary rather than competing, and stating that explicitly dissolves the apparent contradiction between “compute one representative structure” and "screen the whole ensemble." We advocate MLIP sampling as the primary engine—it is the only route that surveys, rather than samples, the $\approx 10^4$ symmetry-inequivalent arrangements of even the 3×3×3 cell (Supplementary Table 2) at tractable cost—with an SQS retained as a cheap reference point for the random-limit average, and a cluster expansion invoked for the specific question of whether Ca and Zr exhibit ordering tendencies that a purely random model would miss. The SQS thus answers the first question and MLIP sampling the second; they are used for different purposes, not as substitutes. The essential caveat, developed in §6.5, is that the MLIP energy ranking of a doped perovskite is trustworthy only after the potential has been fine-tuned on a chemistry-specific DFT dataset, because solid-solution perovskites are under-represented in the public training data.

### 6.2 Multi-Tier DFT Hierarchy: From r²SCAN to HSE06 Refinement

The configurations that survive the machine-learning screen—both the configurational survey of §6.1 and the finite-temperature filter of §6.3—are re-evaluated with density functional theory along the "Jacob's-ladder" hierarchy of §4.2, with each rung matched to the quantity it must deliver. A cheap PBEsol pre-relaxation [62] furnishes starting geometries that the r²SCAN meta-GGA [64] then refines and energy-ranks—the ranking on which the stability ordering of the survivors rests—recovering most of the accuracy of higher rungs at only a few times the GGA cost. Because r²SCAN still underestimates the band gap, the screened hybrid HSE06 [68]—which reproduces the experimental gap of $BaTiO_3$ and, more broadly, the electronic and structural properties of ferroelectric oxides [69]—is used to obtain the gap and, through it, the intrinsic-leakage floor of the insulating layer (§4.2).

The governing design constraint is that HSE06 is computationally severe: the exact-exchange evaluation raises its cost by roughly one-and-a-half to two orders of magnitude over a semilocal functional (Supplementary Table 3), and this penalty scales prohibitively with cell size. Running HSE06 on the 135-atom (and larger) supercells needed to represent the disorder is therefore usually intractable, and any workflow implying otherwise is not credible. In the proposed hierarchy HSE06 appears only as single-point gap evaluations on the small, final down-selected structures, while r²SCAN carries the energetic workload across the broader candidate set. This tiering—cheap functionals for the many, expensive functionals for the few—is precisely what keeps the accuracy requirement of §4.2 compatible with the throughput requirement of §4.1.

### 6.3 Finite-Temperature Stability and Lattice Dynamics — Beyond QHA

A 0 K energetic ranking is not physically meaningful for BZT–BCT, whose ferroelectric rhombohedral, orthorhombic and tetragonal phases lie within a few meV per formula unit—well below the 26.7 meV of thermal energy at body temperature (§4.3). The framework must therefore compute free energies, not static energies, at 310 K. The quasi-harmonic approximation (QHA, via Phonopy [73]) is the natural entry point but is fundamentally inadequate here: it assumes each phonon frequency depends only on volume, and so cannot treat the imaginary soft modes of the high-symmetry cubic reference [74] or the strong anharmonicity that governs the near-MPB region. Far from being a weakness of the approach, this is exactly why an explicitly anharmonic, finite-temperature treatment is required—and why an affordable surrogate for the energy surface is indispensable.

The physically correct tools are anharmonic: MLIP-driven molecular dynamics, the temperature-dependent effective potential (TDEP) [75], and the stochastic self-consistent harmonic approximation (SSCHA) [76], each of which demands $10^3$–$10^6$ energy and force evaluations that only a machine-learning potential can supply at scale. The precedent is direct. The effective-Hamiltonian and DFT-parametrized molecular-dynamics approaches that first reproduced the temperature-driven phase sequence of $BaTiO_3$ [71], [74] have since been generalized by universal perovskite potentials such as UniPero, which recovers the experimental transition sequences of multi-cation solid solutions [86]. Applied to BZT–BCT, this tier yields the Gibbs free energy F(T), the phase populations, and the near-MPB dielectric and elastic softening at 310 K—the very quantities that decide whether a candidate composition remains thermally stable and linear across the arterial-pulse operating window (§3.2). Because this screen is carried out entirely with the potential, it is affordable enough to precede—rather than follow—the density-functional stages, and it therefore occupies tier 2 of the funnel (Table 6).

### 6.4 Property Computation on the Survivors: $d_{33}$ via DFPT and Berry-Phase Polarization

Only at the tip of the funnel are the sensor observables computed, and here the honest limitation of §5.2 is decisive: a machine-learning interatomic potential returns energies, forces and stresses, but not the electronic and polarization quantities that define a piezoelectric transducer. The spontaneous polarization $P_s$ is a Berry-phase property of the occupied wavefunctions, evaluated through the modern theory of polarization [95], [96]; the Born effective charges $Z^*$, the dielectric tensor ε and the piezoelectric tensor $d_{33}$ follow from density-functional perturbation theory (DFPT) [97], which treats atomic displacements, strains and homogeneous electric fields—and their mixed responses—within a single systematic framework [98]. These calculations are expensive and are therefore restricted to the small set of thermally stable, well-insulating survivors emerging from §6.1–6.3.

Two caveats must be stated plainly, because they bound what the framework can legitimately claim. First, the DFPT piezoelectric response is defined for a specific, ordered, fully relaxed cell, whereas the material is a disordered solid solution; a defensible estimate therefore requires computing $d_{33}$ on several representative low-energy configurations drawn from the ensemble of §6.1 and averaging them (ideally Boltzmann-weighted), not reporting a single arrangement. Second, the standard calculation returns the 0 K clamped- and relaxed-ion coefficients, whereas the device operates at 310 K, where the near-MPB softening (§6.3) both enhances and destabilizes the response; the static value is thus a lower bound on, and an imperfect proxy for, the operating $d_{33}$, and the bridge to a dynamic device output requires the multiscale step deferred to §7.1. A promising medium-term route to relieve this DFPT bottleneck is emerging—symmetry-adapted, equivariant networks trained to predict tensorial properties directly, including Born charges, dielectric and polarization response [90], [91], could eventually bring property prediction inside the high-throughput loop. We emphasize, however, that such property models remain specialized and data-hungry, and that no universal potential predicts $d_{33}$ today; keeping this near-term reality distinct from the longer-term goal is what separates a credible perspective from an over-claim.

### 6.5 Data Provenance, Benchmarking and Uncertainty Quantification

The credibility of the whole pipeline rests on the trustworthiness of its machine-learning tier, which is not automatic. Doped, solid-solution perovskites such as BZT–BCT are sparsely represented in the public databases—the Materials Project [54], OQMD [55] and Alexandria [56]—on which universal potentials are trained, so an out-of-the-box model operates outside its domain of validity precisely where quantitative energy ranking is demanded. Two established remedies apply. Active learning selects, on the fly, the configurations for which the potential is least certain and adds their DFT labels to the training set, ensuring that predictions are interpolations rather than extrapolations [79]; Δ-learning instead trains the model to predict the smoother, cheaper-to-learn difference between an inexpensive baseline and the target r²SCAN or HSE reference, raising accuracy for a given data budget [99]. Either route yields the chemistry-specific, fine-tuned potential that §6.1 requires.

Two further safeguards close the loop. Benchmarking against standardized tasks—for example the crystal-stability screening protocol of Matbench Discovery [89]—calibrates the potential's screening performance before it is trusted, while the survivors' computed lattice parameters, Curie temperature and $d_{33}$ must be checked against the available BZT–BCT measurements [18], [19] to anchor the pipeline to experiment. Uncertainty quantification is not optional, and the tools for it are specific rather than generic: the spread of a committee of models gives a calibrated, inexpensive error estimate [100], while an extrapolation grade flags configurations that fall outside the manifold the potential was trained on and can be used to trigger new labelling on the fly [79]. Reporting one of these, and directing targeted DFT spot-checks to the high-uncertainty cases, is now an expected component of any machine-learning-for-materials study. Finally, the framework is deliberately falsifiable—it predicts the 310 K phase populations and transition sequence, the compositional shift of the Curie temperature, and the magnitude and composition-dependence of $d_{33}$, each of which is directly testable against dielectric and piezoelectric experiment. It is this testability, rather than any single computed number, that distinguishes the proposed programme from speculation. One practical corollary deserves stating, because it is what makes the preceding claims auditable rather than merely asserted: a five-tier pipeline of this depth is reproducible only if it is executed under a workflow manager that records the full provenance graph—inputs, code versions, intermediate results and the branch points between tiers—automatically rather than by hand. Infrastructures built for precisely this purpose are now mature [101], and adopting one is what would allow the protocol of Table 6 to be re-run, audited or extended by a third party, and its fine-tuned potential to be traced back to the specific configurations that trained it. The framework, however, stops at the intrinsic single-crystal coefficient; translating that coefficient into a device-level voltage on deforming skin is the multiscale problem taken up in Section 7.

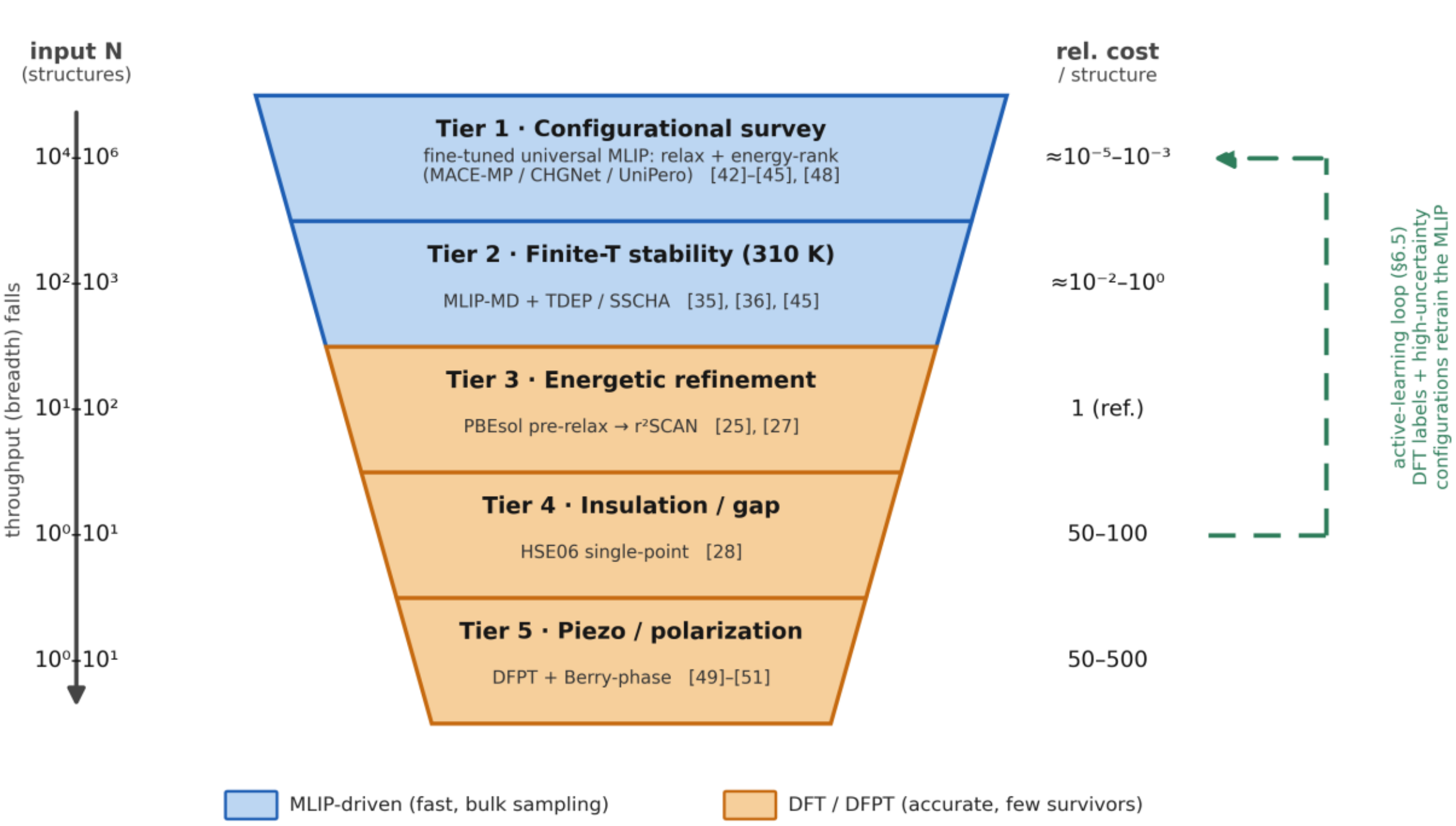

Figure 15. The proposed five-tier screening funnel. A fine-tuned universal MLIP first triages $10^4$–$10^6$ configurations by relaxed energy (tier 1) and then subjects the $10^2$–$10^3$ survivors to finite-temperature MD / TDEP / SSCHA sampling at 310 K (tier 2); only the candidates that remain thermally stable are passed to density functional theory, where r²SCAN refines the energetics (tier 3), HSE06 fixes the band gap and the intrinsic-leakage floor (tier 4), and DFPT with Berry-phase polarization delivers the sensor observables $d_{33}$, $g_{33}$, ε, $Z^*$ and $P_s$ (tier 5). Cost per structure, normalized to one r²SCAN relaxation, rises monotonically by roughly six orders of magnitude from tier 1 to tier 5, while the number of structures falls by five to six orders; throughput is therefore spent where it is cheap and quantum-mechanical accuracy where it is indispensable. The dashed return path is the active-learning loop of §6.5: high-uncertainty configurations, together with the DFT labels generated in tiers 3–5, are fed back to retrain the potential.

*Table 6. The proposed workflow expressed as a staged protocol. Tiers are ordered by increasing cost per structure, so that both machine-learning stages—the configurational survey and the finite-temperature screen—are completed before any density-functional calculation is attempted; the section developing each tier is given in parentheses. "Relative cost per structure" is normalized to one r²SCAN relaxation of the same supercell (tier 3 ≡ 1) and denotes the complete task performed on that structure, not a single energy call: tier 1 is one MLIP relaxation, whereas tier 2 requires $10^5$–$10^6$ force calls per structure and is accordingly two to three orders of magnitude more expensive, while still remaining at or below the cost of a single r²SCAN relaxation. Structure counts are indicative order-of-magnitude targets for a BZT–BCT campaign, not hard limits.*

| Tier / stage | Method & tool | Input N | Primary output | Rel. cost / struct. |
|---|---|---|---|---|
| **1 — Configurational survey (§6.1)** | Fine-tuned universal MLIP: relax + energy-rank (MACE-MP / CHGNet / UniPero) [58], [83]–[86] | $10^4$–$10^6$ | E, F, σ; energy-ranked inequivalent ensemble | ≈$10^{-5}$–$10^{-3}$ |
| **2 — Finite-T stability at 310 K (§6.3)** | MLIP-MD + TDEP / SSCHA [75], [76], [86] | $10^2$–$10^3$ | F(T), phase populations at 310 K | ≈$10^{-2}$–$10^0$ |
| **3 — Energetic refinement (§6.2)** | PBEsol pre-relax → r²SCAN [62], [64] | $10^1$–$10^2$ | ΔE stability ranking of survivors | 1 (reference) |
| **4 — Insulation / gap (§6.2)** | HSE06 single-point [68] | $10^0$–$10^1$ | $E_g$; intrinsic-leakage floor | 50–100 |
| **5 — Piezo / polarization (§6.4)** | DFPT + Berry-phase [95], [97], [98] | $10^0$–$10^1$ | $d_{33}$, $g_{33}$, ε, $Z^*$, $P_s$ | 50–500 |

## 7. Future Perspectives and Clinical Device Integration

The framework of Section 6 terminates in an intrinsic, single-crystal material coefficient—$d_{33}$, $g_{33}$, ε and the spontaneous polarization $P_s$—for the surviving BZT–BCT composition at 310 K. A clinically useful arterial-pulse e-skin, however, is not a bulk single crystal but a thin, flexible ceramic–polymer composite that must convert a 1–10 kPa pressure wave into a readable, self-powered voltage while worn on living skin. Three gaps therefore separate the computed coefficient from a working device: the multiscale leap from a static, atomic-scale polarization to a dynamic device voltage (§7.1); the biological interface of biocompatibility, encapsulation and conformal skin contact (§7.2); and, furthest ahead, the atomistic description of the wet, dynamic skin interface itself (§7.3). We treat these as forward-looking perspectives and mark explicitly where each is near-term engineering and where it is long-horizon speculation.

### 7.1 From Static Polarization to Dynamic Heartbeat Transduction: A Multiscale Bridge

It is tempting to equate the DFT-computed spontaneous polarization or piezoelectric coefficient directly with device performance, but this is a category error: the first-principles calculation returns an intrinsic, 0 K-referenced, single-crystal property (§6.4), whereas the clinical observable is a time-resolved open-circuit voltage $V_{oc}(t)$ produced by a specific composite architecture under a specific pressure waveform. Bridging the two demands a multiscale chain in which each length scale supplies the constitutive input for the next (Figure 16); no single method spans it, and DFT alone certainly does not.

The chain has three stages. (i) At the atomic scale, DFT/DFPT delivers the single-crystal constitutive tensors of the survivor composition—the piezoelectric d, dielectric ε and elastic C tensors (§6.4)—which are the material constants, not the device output. (ii) At the microstructural scale, these single-crystal constants are homogenized into effective composite coefficients: because the active ceramic is dispersed in a compliant polymer with a specific connectivity (0–3, 1–3 or 2–2 in Newnham's classic scheme [38]) and porosity, effective-medium and micromechanical models—the mean-field averaging scheme of Mori and Tanaka [102], extended to the coupled electroelastic problem of a piezoelectric inclusion in a dielectric matrix by Dunn and Taya [103]—convert the intrinsic $d_{33}$, ε and ceramic volume fraction into the composite's effective $d_{33,\mathrm{eff}}$ and $g_{33,\mathrm{eff}}$. This is precisely the stage at which the sensitivity–flexibility trade-off of §2.4 becomes quantitative—and where the counter-intuitive gain in $g_{33}$ obtained by embedding a high-permittivity ceramic in a low-permittivity matrix (§2.1) is captured. (iii) At the device scale, a finite-element electromechanical model of the actual geometry—film thickness, electrode layout, substrate and packaging—driven by the measured arterial pressure waveform yields $V_{oc}(t)$, together with the response time, the linearity across 1–10 kPa, and the fidelity of features such as the dicrotic notch; the linear-piezoelectric constitutive framework and the distributed-parameter and finite-element machinery for exactly this class of problem are well established for piezoelectric harvesters and sensors [104].

Framed this way, the computational programme of Sections 4–6 is not a device simulator but the supplier of trustworthy, temperature-correct material constants to the top of this bridge; the homogenization and finite-element stages are mature, experimentally validated engineering tools, and the predicted $V_{oc}(t)$ can be checked directly against the wrist-pulse and heart-sound waveforms already reported for fabricated lead-free devices [7], [21], and ultimately against the validation protocols that govern cuffless measurement [52], [53]. Making the multiscale hand-off explicit—rather than implying that a device voltage falls straight out of a Berry-phase polarization—is what keeps this perspective physically honest.

Figure 7.1 Multiscale bridge from first-principles constants to dynamic device voltage

*DFT supplies the material constants · homogenization builds the composite · FEM produces the device voltage*

**Atomic scale**
*DFT / DFPT · Berry phase (single crystal)*

d, ε, C

**Microstructure**
*homogenization + porosity 0–3 / 1–3 / 2–2 connectivity*

d₃₃,eff

**Device**
*finite-element model pressure 1–10 kPa*

output
single-crystal tensors
d , ε , C , Ps

output
effective composite
d₃₃,eff , g₃₃,eff

output
Voc(t) · cardiac cycle

~Å – nm
~µm
~mm (device)
*length scale*

Figure 16. The multiscale bridge from first-principles material constants to a dynamic device voltage. DFT/DFPT supplies the intrinsic single-crystal tensors (d, ε, C and the polarization $P_s$); micromechanical homogenization converts these—together with the ceramic volume fraction, the connectivity (0–3, 1–3 or 2–2) and the porosity—into the effective composite coefficients ($d_{33,\text{eff}}$, $g_{33,\text{eff}}$); and a finite-element electromechanical model of the device geometry, driven by the 1–10 kPa arterial pressure wave, yields the time-resolved open-circuit voltage $V_{oc}$(t), including features such as the dicrotic notch. Each stage supplies the constitutive input for the next, and no single method spans the ångström-to-millimetre range.

## 7.2 Biocompatibility, Encapsulation and the Epidermal Interface

The biosafety mandate of §1.2 motivated the move to lead-free chemistry, but “lead-free” is a necessary, not a sufficient, condition for a safe worn device; three interfacial requirements remain. First, intrinsic biocompatibility: $BaTiO_3$ is among the best-tolerated piezoelectric oxides, showing good compatibility with fibroblasts and neuronal cells [17], and it belongs to the broader class of piezoelectric biomaterials now being engineered for sensors and actuators in direct contact with tissue [105]; a lead-free piezoceramic thin film in flexible form has already passed a comprehensive biocompatibility panel [48]. That compatibility is nonetheless dose-, size- and cell-type-dependent, and for the BZT–BCT composition the relevant precaution shifts from $Pb^{2+}$ toxicity to controlling any release of $Ba^{2+}$ or dopant ions—an argument for barrier encapsulation rather than for reliance on the intrinsic chemistry alone.

Second, encapsulation. A skin-worn transducer must be sealed against outward ion leaching and against inward ingress of sweat and biofluids, yet remain thin and compliant enough not to attenuate the very 1–10 kPa signal it is meant to sense. Conformal barrier films such as parylene-C and elastomer/inorganic multilayers provide exactly this combination of biocompatibility, moisture barrier and flexibility, and the encapsulation of flexible bioelectronic implants is now a developed engineering discipline with quantified lifetimes and characterized failure modes such as adhesion loss and moisture permeation [51]. Crucially, it is the encapsulation and electrode design—not the active layer's intrinsic band gap (§4.2)—that actually govern electrical safety at the skin, which closes the loop on the leakage-and-insulation argument made earlier: the band gap sets the material's intrinsic insulation quality, whereas patient safety is a device-integration property.

Third, the epidermal interface. Faithful pulse capture requires a mechanical-impedance match to skin (in vivo indentation modulus of order 10–100 kPa, §1.1 [9]) and intimate, motion-tolerant conformal contact on a curved, deforming, perspiring surface; the epidermal-electronics design principles that match a device's thickness, effective modulus and bending stiffness to those of the epidermis [106] are the template for achieving this without the motion artefacts that degrade rigid transducers. These are engineering frontiers that the computational framework of Section 6 informs only indirectly—by identifying a biocompatible, high-sensitivity active material—while the encapsulation and the skin interface remain device-integration tasks to be solved experimentally.

### 7.3 Open Challenge: Machine-Learning Potentials for the Wet, Dynamic Skin Interface

The most distant frontier returns the problem to the computational method itself. The entire framework of Section 6—and every universal MLIP it relies on—describes the dry, crystalline material and its bulk finite-temperature behaviour. The true operating environment is neither dry nor static: it is a hydrated, ionic, chemically active skin interface, bathed in sweat electrolytes and subject to biofouling, in which the surface polarization, the charge screening and the long-term chemical stability may all depart from the bulk prediction. Whether the giant, MPB-derived response survives at such an interface over months of continuous wear is, at present, an open question that bulk computation cannot answer.

Addressing it would require machine-learning potentials trained for solid–liquid interfaces—oxide surfaces in contact with water and dissolved ions—including the reactive and electrochemical processes and the long-timescale dynamics that such systems entail. This is an active and fast-moving area: data-driven, active-learning protocols now make it feasible to construct MLIPs for complex aqueous and interfacial systems, including water structuring and wetting at surfaces [107]. Oxide–water interfaces in particular have become a recognized application domain for such potentials, and the methodological requirements they impose—a reactive description of surface hydroxylation and proton transfer, adequate treatment of long-range electrostatics, and sampling long enough to converge the structure of the interfacial region—are by now reasonably well mapped [108]. The same equivariant architectures that power the bulk framework [77], [81], [83]–[86], [90], [91] are being extended to these environments. But a quantitatively reliable, doped-perovskite/sweat-interface potential—one that could predict how the operating $d_{33}$ and the stability of a specific BZT–BCT surface evolve in a physiological electrolyte—remains well beyond current turnkey capability. We therefore present this not as a near-term deliverable but as a clearly delimited open challenge: it marks the boundary of the framework, whose validated domain is the bulk material at body temperature, and it defines a concrete direction for the coming decade of work.

## 8. Conclusions

Continuous, non-invasive arterial-pulse monitoring calls for a flexible, self-powered transducer, and among lead-free piezoelectrics the $BaTiO_3$-based BZT–BCT solid solution uniquely couples near-soft-PZT sensitivity with the biocompatibility that skin contact demands. We have argued that the decisive obstacle to computationally discovering and optimizing such a material is not a single missing calculation but configurational complexity: the astronomically large space of inequivalent dopant arrangements in a multi-component solid solution, which conventional 0 K density functional theory can neither enumerate at tractable cost nor evaluate at the 310 K temperature where the device must operate.

The resolution we have advanced is a staged quantum–AI framework built on equivariant machine-learning interatomic potentials. A fine-tuned universal potential surveys—rather than samples—the configurational ensemble and, through finite-temperature molecular dynamics, delivers the body-temperature thermodynamics that the quasi-harmonic approximation cannot, while a multi-tier DFT hierarchy ($r^2$SCAN energetics with HSE06 gaps reserved for the final candidates) and DFPT/Berry-phase property evaluation act only on the small set of survivors. Cast as a screening funnel, the workflow spends computational throughput where it is cheap and quantum-mechanical accuracy where it is indispensable, and it maps each of the three walls of conventional DFT—configurational explosion, band-gap error, and the 0 K limitation—onto a distinct, tractable tier.

Two points determine whether such a programme is credible rather than promotional, and we have kept both explicit throughout. First, the division of labour must be stated honestly: a machine-learning interatomic potential predicts energies, forces and stresses—and hence structural relaxation, thermodynamic stability and finite-temperature behaviour—but not the electronic and polarization observables (the band gap $E_g$, the spontaneous polarization $P_s$, the Born effective charges $Z^*$, the dielectric tensor $\varepsilon$, and therefore the piezoelectric coefficient $d_{33}$) that define a piezoelectric sensor; these remain the province of DFT/DFPT or of dedicated, still-maturing tensorial-property networks.

Conflating energy prediction with property prediction is the most common overstatement in this field, and avoiding it is what separates a defensible perspective from hype. Second, the computational pipeline must be anchored to experiment: its predicted stabilities, transition temperatures and piezoelectric coefficients are falsifiable against measured data, and—because the highest-sensitivity BZT–BCT composition has still not been realized as a flexible pulse sensor—the framework's ultimate value rests on closing the loop with fabrication and device testing, through the multiscale bridge that converts an intrinsic single-crystal coefficient into a dynamic device voltage.

Realizing this vision requires advances that are already under way but not yet turnkey: chemistry-specific fine-tuning, active learning and uncertainty quantification to make universal potentials trustworthy for the under-represented chemistry of doped solid solutions; equivariant models that predict tensorial properties directly, so that $d_{33}$ screening can eventually enter the high-throughput loop; and, further ahead, interfacial potentials able to describe the wet, dynamic skin environment in which the device must finally operate. Should these mature, the computational discovery of lead-free piezoelectric perovskites would shift from the biased evaluation of a handful of hand-built structures to a systematic, temperature-correct survey of the full configurational space—accelerating the path to a clinically viable, biocompatible arterial-pulse e-skin, and offering a template that transfers directly to the broader class of functional, disordered perovskite oxides.

## Data availability

All data discussed in this article are available within the article, its Supplementary Information, and the cited primary literature. The numerical values underlying every original figure — including the digitized literature values compiled for Figures 3, 4, 7, 9, 12 and 13 and the tabulated figures of merit of Tables 1 and 2, and Supplementary Tables 1–3 — are provided as machine-readable CSV files in the Supplementary Information. The combinatorial counts, achieved compositions, configurational entropies and thermal-energy scales reported in Section 4 and Supplementary Table 2 are exact derived quantities and are reproducible from the formulae given in the text. No new experimental measurements or first-principles datasets were generated for this Review.

## Code availability

The Python (matplotlib) scripts used to generate all original figures in this article are provided in the Supplementary Information and are additionally available from the corresponding author on reasonable request. No custom simulation software was developed for this Review; the computational methods discussed — special quasirandom structures, cluster expansion, density functional theory and density-functional perturbation theory, lattice-dynamics sampling, and machine-learning interatomic potentials — are implemented in the publicly available packages described in the cited primary references.

## Acknowledgements

[To be completed by the authors: funding sources, grant numbers and institutional support.]

## Author contributions

[To be completed by the authors: e.g. A.A. conceived the scope and wrote the manuscript; B.B. prepared the figures and the comparative tables; C.C. supervised the work. All authors discussed the content and approved the final manuscript.]

## Competing interests

The authors declare no competing interests.

## References

***Note.*** Unified reference list for Sections 1–8, numbered in a single sequence in order of first appearance in the text. Derived or computed quantities (the combinatorial counts, achieved compositions, configurational entropy and thermal-energy scales of Section 4) are exact results reproducible from the formulae given, and are intentionally not cited. Entries without a DOI are standards, statutes or monographs, for which none is issued.